\documentclass[a4paper,11pt]{article}
\pdfoutput=1 % if your are submitting a pdflatex (i.e. if you have
\usepackage{jcappub} % for details on the use of the package, please
\usepackage[dvipsnames]{xcolor}
\usepackage[T1]{fontenc} % if needed
\usepackage{multirow}
\usepackage{hyperref}
\usepackage{silence}
\title{\boldmath Cosmological Remapping for Efficient Generation of 21 cm Intensity Mapping Mocks}

\author[a]{Rahima Mokeddem,\note{Corresponding author.}}
\author[a,b]{Bruno B. Bizarria,}
\author[c,d]{Jiajun Zhang,}
\author[e,f]{W.S. Hipólito-Ricaldi,}
\author[a]{Carlos Alexandre Wuensche,}
\author[g,h,i]{Elcio Abdalla,}
\author[g,j,k,l]{Filipe B. Abdalla,}
\author[m]{Amilcar R. Queiroz,}
\author[a,n,o]{Thyrso Villela,}
\author[p,q]{Bin Wang,}
\author[j,q,r]{Chang Feng,}
\author[s]{Edmar C. Gurjão,}
\author[i,j]{Alessandro Marins}

\affiliation[a]{Instituto Nacional de Pesquisas Espaciais,Divisao de Astrofísica,\\Av. dos Astronautas, 1758, CEP 12227-010, São José dos Campos, SP, Brazil}

\affiliation[b]{Jodrell Bank Centre for Astrophysics, Department of Physics \& Astronomy, The University of Manchester, Manchester M13 9PL, UK}

\affiliation[c]{Shanghai Astronomical Observatory, Chinese Academy of Sciences, Shanghai, China}

\affiliation[d]{State Key Laboratory of Radio Astronomy and Technology, 20A Datun Road, Chaoyang District, Beĳing, 100101, China}

\affiliation[e]{Departamento de Ci\^encias Naturais, CEUNES, Universidade Federal do Esp\'{\i}rito Santo, Rodovia BR 101 Norte, km. 60, CEP 29.932-540, S\~ao Mateus, ES, Brazil.}

\affiliation[f]{Núcleo Cosmo-UFES, Universidade Federal do Espírito Santo, Av. Fernando Ferrari, 540, CEP 29.075-910, Vitória, ES, Brazil.}

\affiliation[g]{Instituto de Física, Universidade de São Paulo,\\ C.P. 66.318, CEP 05315-970, São Paulo, Brazil}

\affiliation[h]{Universidade Estadual da Paraíba,\\ Rua Baraúnas, 351, Bairro Universitário, Campina Grande, Brazil}

\affiliation[i]{Departamento de Física, Centro de Ciências Exatas e da Natureza, Universidade Federal da Paraíba,\\ CEP 58059-970, João Pessoa, Brazil}

\affiliation[j]{Department of Astronomy, School of Physical Sciences, University of Science and Technology of China,\\  Hefei, Anhui 230026, China}

\affiliation[k]{School of Astronomy and Space Science, University of Science and Technology of China,\\ Hefei,  Anhui 230026, China}

\affiliation[l]{Department of Physics and Electronics, Rhodes University,\\ PO Box 94, Grahamstown, 6140, South Africa}

\affiliation[m]{Unidade Acadêmica de Física, Univ. Federal de Campina Grande,\\ R. Aprígio Veloso, 58429-900 - Campina Grande, Brazil}

\affiliation[n] {Centro de Gest\~ao e Estudos Estrat\'egicos SCS Qd 9, Lote C, Torre C S/N Salas 401 a 405, 70308-200 - Bras\'ilia, DF, Brazil}

\affiliation[o] {Instituto de F\'{i}sica, Universidade de Bras\'{i}lia, Campus Universit\'ario Darcy Ribeiro, 70910-900 - Bras\'{i}lia, DF, Brazil }

\affiliation[p]{Center for Gravitation and Cosmology, Yangzhou University,\\ Yangzhou 224009, China}

\affiliation[q]{School of Aeronautics and Astronautics, Shanghai Jiao Tong University,\\ Shanghai 200240, China}

\affiliation[r]{CAS Key Laboratory for Research in Galaxies and Cosmology, University of Science and Technology of China,\\ Hefei, Anhui 230026, China}

\affiliation[s]{Departamento de Engenharia Elétrica, Univ. Federal de Campina Grande,\\ R. Aprígio Veloso, 58429-900 - Campina Grande,
Brazil}

\emailAdd{rahima.mookeddem@inpe.br}
\abstract{We present a novel application of cosmological rescaling, or "remapping," to generate 21 cm intensity mapping mocks for different cosmologies. The remapping method allows for computationally efficient generation of N-body catalogs by rescaling existing simulations. In this work, we employ the remapping method to construct dark matter halo catalogs, starting from the Horizon Run 4 simulation with WMAP5 cosmology, and apply it to different target cosmologies, including WMAP7, Planck18 and Chevallier-Polarski-Linder (CPL) models.  These catalogs are then used to simulate 21 cm intensity maps. We use the halo occupation distribution (HOD) method to populate halos with neutral hydrogen (HI) and derive 21 cm brightness temperature maps. Our results demonstrate the effectiveness of the remapping approach in generating cosmological simulations for large-scale structure studies, offering an alternative for testing observational data pipelines and performing cosmological parameter forecasts without the need for computationally expensive full N-body simulations. We also analyze the precision and limitations of the remapping, in light of the rescaling parameters $s$ and $s_m$, as well as the effects of the halo mass and box size thresholds.
}

\begin{document}
\maketitle
\flushbottom

\section{Introduction}
\label{sec:intro}

In the last decades, the science of cosmology has been facing a period of fast advancement, mainly due to a set of cosmological observables surveyed with increasingly better precision, such as the temperature fluctuations and polarization of the Cosmic Microwave Background (CMB) (e.g., \citep{collaboration2020planck,Smoot1992,Komatsu2009}), Supernovae Type Ia \citep{riess1998observational,perlmutter1998cosmology}, Baryon Acoustic Oscillations (BAO) (e.g.,\citep{Einsenstein2005,Anderson2014,troster2019cosmology}), and cosmic shear \citep{secco2022dark, Kuijken2025}. Recent results from experiments like the Atacama Cosmology Telescope (ACT) \citep{Choi2020}, the Euclid mission \citep{EuclidCollaboration2022}, and the Dark Energy Spectroscopic Instrument (DESI) \citep{DESICollaboration2023} have significantly enhanced our understanding of cosmological parameters and the late-time universe. In particular, DESI's initial data release provides unprecedented constraints on dark energy and the expansion history of the universe, highlighting the importance of robust methods for analyzing complex cosmological models, such as those with dynamical dark energy parameterized by the CPL form \citep{Chevallier2001,Linder2003}. Furthermore, gravitational lensing, including cosmic shear \citep{secco2022dark} and galaxy-galaxy lensing (e.g., \citep{BartelmannSchneider2001}), probes the distribution of dark matter and the growth of structure. The study of galaxy clusters, the most massive gravitationally bound structures in the universe, provides crucial insights into cosmology through their abundance, spatial distribution, and internal properties (e.g., \citep{Allen2011, Voit2005}). Other important probes include redshift-space distortions (e.g., \citep{Hamilton1998}), which allow us to measure the peculiar velocities of galaxies and constrain the growth rate of structure, and the Lyman-alpha forest (e.g., \citep{McDonald2006,Ramirez-Perez2024}), which probes the intergalactic medium at high redshifts.

One additional observable is the so-called 21 cm hyperfine transition of electrons in neutral hydrogen (HI).
The 21 cm signal arises from the transition between two levels of the hyperfine splitting of the ground state of HI, caused by the spin-spin interaction between the proton and electron.\\
The $21 cm$ signal can be used to trace matter density across cosmic eras, mapping large areas in the sky using a technique known as Intensity Mapping. It provides an alternative, as well as complementary, cosmic observable that is highly useful for cosmological analyses. In fact, recent intensity mapping results from the MeerKAT telescope, as presented by \citep{Cunnington_2022, carucci2024}, represent the state-of-the-art in single-dish intensity mapping, demonstrating the potential of this technique for large-scale structure studies. In recent years, several experiments have been proposed to measure the $21 cm$ signal through intensity mapping, for example, BINGO \citep{Wuensche2019, Abdalla2022, Wuensche2022}, CHIME \citep{2022chime}, Tianlai \citep{Chen2012}, FAST \citep{Bigot-Sazy2016, Smoot2017, Hu2020}, and the SKA \citep{SKA2018}.

To extract information from the observational data, it is necessary to compare them with theoretical predictions.
Usually, astronomical projects perform many different forecasting processes, generating simulations to assess the desired instrument performance and check for problems in the design, long before commissioning and observation campaigns take place. Thus, the generation of 21 cm mocks is essential because it enables the testing of the data analysis pipeline and the assessment of the reliability of cosmological parameter constraints. It also provides a means to quantify the range of errors and uncertainties in those constraints and to perform forecasting for specific observational projects, offering valuable insights for their design and optimization.

However, constructing 21 cm mocks depends critically on accurately modeling the distribution of neutral hydrogen (HI) mass in the universe, a key ingredient that must be quantified. Several approaches exist to achieve this: \citep{Villaescusa_Navarro_2018} derive HI properties directly from hydrodynamic simulations; \citep{Koda2016} use halo catalogs generated by a fast simulation method (COLA-HALO); \citep{Alonso2014} and \citep{Xavier2016} adopt a lognormal dark matter density distribution; \citep{Zhang2019} utilize galaxy catalogs based on N-body simulations; and \cite{seehars2016simulating}, \citep{Asorey2020}, and \citep{Zhang_2022} rely on dark matter halo distributions derived from N-body simulations.

Methods based on N-body simulations face a significant challenge: they are computationally demanding and limited in flexibility. Most available N-body simulations are tailored to specific cosmological models and rely on highly constrained parameter sets. If one wishes to explore a wide range of cosmological parameter space, including variations like neutrinos, curvature, or warm dark matter, or test alternative cosmological models, new simulations must be run for each set of parameters. For each case, the matter distribution must be recalculated, and halo catalogs must be generated anew. This requirement makes the process computationally inefficient, time-consuming, and impractical for large-scale explorations of parameter space. To address this issue, alternative rescaling methods that reuse existing simulations have been proposed, allowing the exploration of different cosmologies without running new simulations.

One such approach is the 'cosmology rescaling' or 'remapping' technique, originally proposed by \citep{Angulo2010}. This method exploits the near-universality of the mass function \citep{Sheth1999, Tinker2008} to transform an N-body particle distribution into one that approximates a simulation with a different cosmology. We define the original cosmology as the one simulated and the target as the one we aim to approximate. By rescaling lengths, masses, and times, and selecting snapshots that match the %linear density
fluctuation amplitude of the target cosmology, this method eliminates the need for rerunning simulations from scratch, enabling the exploration of a broader parameter space. Later, \citep{Mead2014a} refined the technique, making it self-contained and directly applicable to pre-existing halo catalogs without the need for additional information. The remapping method has been successfully applied in various contexts, including redshift space, modified gravity models, weak galaxy-galaxy lensing, and cosmologies with massive neutrinos \citep{Mead2014b, Mead2015, Renneby2018, Zennaro2019}.

This work investigates the potential of applying the remapping technique to dark matter halo catalogs (DMHC hereafter) as a robust and efficient method for generating 21 cm brightness temperature mocks. Building upon our initial exploration of this remapping approach resulting in WMAP7 21cm mocks \citep{Mokeddem:2023zaa}, the ultimate goal is to produce mocks relevant for the BINGO radiotelescope's~\citep{Wuensche2022} frequency band and redshift range, we generated 21 cm maps across a broader spectrum of frequencies, dictated by the available redshift outputs of the N-body simulations. This approach allows for a comprehensive assessment of the remapping technique's validity across different scales.The consistent agreement observed across this wider frequency range provides strong evidence for the reliability and applicability of the remapping technique, including its potential for generating accurate 21 cm mocks for the BINGO experiment. To populate these dark matter halos with neutral hydrogen (HI) and subsequently generate 21 cm signals, we employed the HOD method proposed by \cite{Zhang_2022}, highlighting its potential for future large-scale structure studies. Furthermore, we analyze the precision and limitations of the remapping, in light of the rescaling parameters s and $s_m$, as well as the effects of the halo mass and box size thresholds. Accordingly, our goal is not merely to demonstrate an application of cosmological rescaling, but to thoroughly investigate its limitations.

This paper is organized as follows: In Section \ref{sec:methodology}, we briefly describe the remapping methodology and the generation of 21~cm mock based on the HOD approach. Section \ref{sec:data} oulines the N-body simulation used in this work. Our findings are presented in Section \ref{sec:results}. The analysis of remapping precision is provided in Appendi~\ref{Remapping precision}, where we establish limits on the remapping parameters to ensure the reliability of the technique.
%%%%%%%%%%%%%%%%%%%%%%%%%%%%%%%%%%%%%%%%%%%%%%%%%%%%5
\section{Methodology}\label{sec:methodology}

%%%%%%%%%%%%%%%%%%%%%%%%%%%%%%%%%%%%%%%%
Our methodology for constructing 21 cm mocks in different cosmologies consists of two main steps. First, we use the DMHC remapping method \citep{Mead2014a} to generate halo catalogs for any target cosmology from a single N-body simulation. Second, we use these outputs to create 21 cm intensity map mocks, following the approach adopted by the BINGO project \citep{Zhang_2022}. Both steps are described in the following subsections.
%%%%%%%%%%%%%%%%%%%%%%%%%%%%%%%%%%%%%%%%
\subsection{Remapping Dark Matter Halo Catalogs}
%%%%%%%%%%%%%%%%%%%%%%%%%%%%%%%%%%%%%%%%
Cosmological rescaling, or remapping, is a technique that converts a catalog generated from an N-body simulation in one cosmology (\textbf{the original cosmology}) into a catalog that corresponds to a different cosmology (\textbf{the target cosmology}). This method was first introduced by \citep{Angulo2010} and later extended by \citep{Mead2014a} to be fully self-contained, requiring no additional information beyond what is available in the halo catalog of the original cosmology. The procedure initially involves two key steps:
\begin{itemize}
    \item Rescaling the time and length scales so that the halo mass function of the original simulation closely matches that of the target cosmology.
    \item Adjusting the particle positions in the new cosmology to reproduce the clustering patterns expected in the target cosmology.
\end{itemize}

In the first step, remapping modifies the redshifts and length scales in the original cosmology to approximate the conditions of a target cosmology at a different redshift, over the same mass range. This is achieved by aligning their halo mass functions, leveraging the near-universality as demonstrated by \citep{Sheth1999} and \citep{Tinker2008}. 

In the framework of Press-Schechter theory \citep{PressSchechter1974}, the halo mass function estimates the number density $n$ of collapsed objects as a function of their mass $M$, and is expressed as:
 \begin{equation} \label{mass_func}
 \frac{dn}{dM} = \sqrt{\frac{2}{\pi}}\frac{\bar{\rho}}{M}\frac{\delta_c}{\sigma^2}\frac{d\sigma}{dM}e^{-\frac{\delta_c^2}{2\sigma^2}}  \,,
 \end{equation}
where $\bar{\rho}$ represents the average matter density,   $\delta_c$ the critical overdensity for collapse and $\sigma$ the smoothed linear variance, given by
\begin{equation} \label{linearvariance}
    \sigma^2(R,z) = \int^{\infty} _{0} \Delta^2_{\text{lin}}(k,z) T^2(kR) d\ln k,
\end{equation}
with $T(kR)$ being  the Fourier transform of a top-hat filter to a radius $R$, 
\begin{equation}
    T(kR) = \frac{3}{kR^3} \left[\sin(kR) - kR \cos(kR)\right] \,,
\end{equation}
in which $R$ stands for the radius containing a mass $M$ in a homogeneous universe i.e $ M=\frac{4\pi}{3} R\bar{\rho} $.

On the other hand the dimensionless linear matter power spectrum in a volume $V$ is given by
\begin{equation}
    \Delta^2_{\text{lin}} (k,z) = 4 \pi V \left(\frac{k}{2\pi}\right)^3 P(k,z) \,,
\end{equation}
where $P(k,z)$ is the power spectrum. Since the mass function in Eq.(\ref{mass_func})  primarily depends  on the power spectrum through the linear variance (\ref{linearvariance}), and the power spectrum
is continuously curved in cosmological models, an appropriate rescaling in redshift and length units can align variances of distinct cosmologies near the nonlinear scale. 
The alignment follows from minimizing the root-mean-square difference  ($\delta_{\text{rms}}$) 
  in linear variance between cosmologies, derived from their halo mass functions \citep{Angulo2010,Mead2014a}.
\begin{equation} \label{rms}
    \delta^2_{\text{rms}} (s,z|z') = \frac{1}{\ln (R_2' / R_1 ')} \int^{R_2'} _ {R_1'} \frac{dR}{R}\left[1-\frac{\sigma(R/s ,z)}{\sigma'(R ,z')}\right]^2 \,.
\end{equation}   

In this work, quantities for the target cosmology are denoted with primes, while those for the original cosmology remain unprimed. By minimizing Eq. (\ref{rms}) for a given original cosmology at redshift $z$, the target redshift $z'$ and scaling length parameter $s$ can be determined  and used to rescale time and lengths units, or vice versa: fixing $z'$, one can solve for $z$ and $s$. The scaled radii are then $R'_1= s\,R_1$ and $R'_2$ = $s\,R_2$, where $R_1$ and $R_2$ correspond to the least and most
massive halos in the original catalogue. The target simulation box size is rescaled as  $L' = s\,L$, implying rescaling wavenumbers in the Fourier space as $k'=s^{-1}k$ and halo masses  as  $M' = s_m M$, where  $s_m = s^3 \Omega '_m/\Omega_m$, with $\Omega_m$ being the  matter density parameter.

 The second phase of the remapping method involves shifting individual halo positions in the rescaled simulation to match the large-scale clustering of the target cosmology.  To do this,  the displacement field $\textbf{f}$ is introduced, moving  particles from their initial Lagrangian positions $\textbf{q}$ to their Eulerian positions $\textbf{x}$ as
\begin{equation} 
    \textbf{x} = \textbf{q} + \textbf{f},    
\end{equation}
for the original rescaled simulation. The displacement field at linear order is given by the Zel'dovich Approximation (ZA) \citep{Zeldovich1970}:
\begin{equation} 
    \label{delta} \delta = - \nabla \cdot \textbf{f} \,, \quad \implies \quad  \textbf{f}_\textbf{k} = -i \frac{\delta_\textbf{k}}{k^2} \textbf{k}.  
\end{equation}
where $\delta$ is the matter overdensity. If the initial conditions are unchanged, halos in the original simulation are shifted from positions $\textbf{x}$ to new positions $\textbf{x}'$ in the target simulation according to:
\begin{equation} 
    \label{shift} \textbf{x}' = \textbf{x} + \delta \textbf{f},   
\end{equation}
with
\begin{equation} 
    \delta \textbf{f}_{\textbf{k}'} = \left[\sqrt{\frac{\Delta'^2_\text{lin}(k',z')}{\Delta^2_\text{lin}(sk',z)}} - 1 \right] \textbf{f}_\textbf{k'}, 
\end{equation}
$\textbf{f}_\textbf{k'}$ is  the displacement field in rescaled original simulation. 

\begin{figure}%[htbp]
    \centering
    \includegraphics[height=8cm, width=11cm]{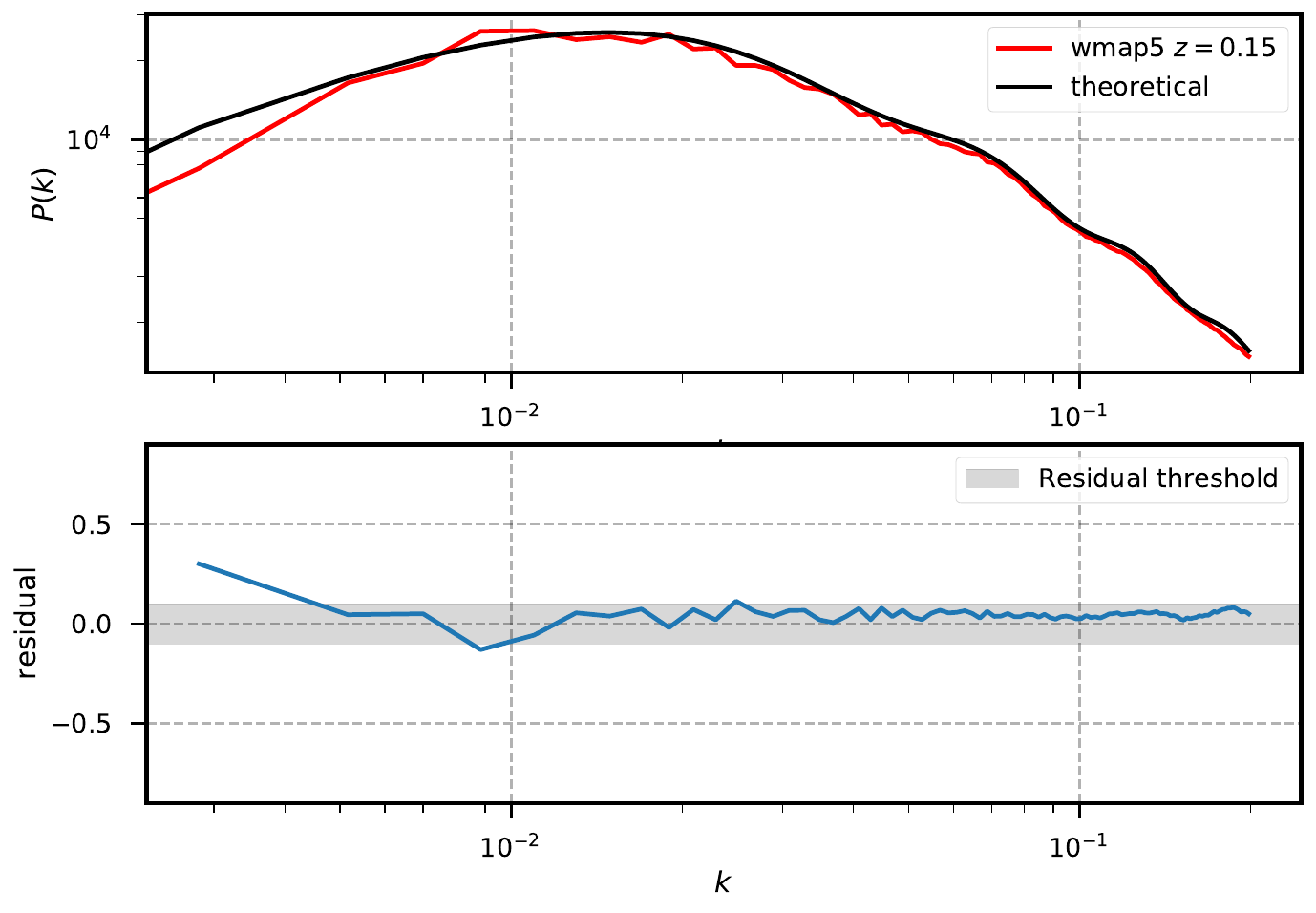}
    \caption{ Comparison of the matter power spectrum P(k) from a WMAP5-based simulation with the corresponding theoretical prediction. The residuals, staying within the $\pm 10\%$ aded region, demonstrate excellent agreement between the simulation and theoretical expectation for the debiased halo power spectrum.}
    \label{fig:hr4_wmap5}
\end{figure}

In the extended remapping method \citep{Mead2014a}, the displacement field $\textbf{f}$ is derived from the density of matter $\delta$, inferred directly from the halo catalogs via the density of the halo number $\delta_\text{H}$. Since halos are biased tracers of the mass distribution, $\delta_\text{H}$ relates to $\delta$ through a bias factor $b$:
\begin{equation} 
    \delta_\text{H} = b \,\delta. 
\end{equation}
%An accurate determination of halo bias is crucial for correctly modeling the relationship between the distribution of dark matter and the observable distribution of galaxies.  
For this study, the halo bias $b$ was calculated using the mass function presented by \citep{Tinker2008} and implemented in the Colossus python package. \citep{Diemer_2018}. %  which provided a robust and efficient framework for computing cosmological quantities. 
It should be noted that there are other approaches for calculating $b$; for example, \citep{Mead2014a} employed a method inspired by \citep{Sheth1999}. Additionally, while a more precise computation may account for the halo mass dependence of the bias, treating $b$ as roughly constant suffices for the precision required in this work.  Figure~\ref{fig:hr4_wmap5} compares the theoretical matter power spectrum from Cosmic Linear Anisotropy Solving System (CLASS) \citep{class} with the halo power spectrum from the Horizon Run 4 simulation (see below), debiased by the factor $b$. At the scales of interest, the residual is always below $10\%$.

Additionally,  a nonlinear scale ${R}'_{nl}$ is defined at each redshift  in the target comology such that  $ \sigma'(R'_\text{nl}, z') = 1$, beyond which fluctuations are considered nonlinear. The corresponding wavenumber $k'_{nl} = {R'}_{nl}^{-1}$ sets a linearity limit for the rescaled modes $k' = s\,k$, defining the scale
at which displacement and density fields transition from the linear to the nonlinear regime. This scale also aids in linearizing the matter over-density field by applying a Gaussian filter with a width set by the nonlinear scale, and in correcting the displacement fields to have the desired variance \citep{Mead2014a} 
\begin{equation} 
    \textbf{f} \rightarrow \textbf{f} \frac{\sigma_f(R_\text{nl})}{\sqrt{\text{Var}(|\textbf{f}|)}},    
\end{equation}
where $\text{Var}(|\textbf{f}|)$ represents the variance of $\textbf{f}$ and
\begin{equation} 
    \sigma_f^2(R_\text{nl}) = \frac{1}{3} \int^{\infty}_{k_\text{box}} \frac{e^{-k^2 R_\text{nl}^2} \Delta^2_\text{lin}(k)}{k^2}  d \ln k\,.     
\end{equation}
where $k_\text{box} = 2\pi / L$ and $L$ is the box length. 

The method described here enables the remapping to a target cosmology from an N-body simulation halo catalog, providing inputs for the next step. It's worth mentioning that a detailed analysis of the remapping precision, including the limitations on the rescaling parameters, is presented in Appendix~\ref{Remapping precision}.

To summarize the remapping process steps: 
\begin{itemize}
    \item Start with a halo catalog from an N-body simulation of the original cosmology and a desired target redshift $z'$.
    \item Using equation~\ref{rms}, determine the scaling parameters ($s$ and $s_m$) and the original redshift ($z$) that will be used for the remapping.
    \item Rescale the properties of the halos in the original catalog: The positions of the halos in the original simulation box are scaled by the factor $s$. The new box size becomes $L'=sL$. Consequently, wavenumbers in Fourier space are rescaled as $k' = s^{-1} k$;
    \item Rescale the halo masses: The masses of the halos in the original catalog are rescaled by the mass scaling parameter $s_m$, so that $M' = s_m M$;
    \item Calculate the displacement field from the halo catalog of the rescaled original cosmology and apply the shift to the rescaled halo positions;
    \item The resulting catalog with rescaled masses and shifted positions now approximates a halo catalog for the target cosmology at redshift $z'$.
\end{itemize}

%%%%%%%%%%%%%%%%%%%%%%%%%%%%%%%%%%%%
\subsection{21 cm Mock Generation}\label{21cm}
%%%%%%%%%%%%%%%%%%%%%%%%%%%%%%%%%%%%

This step begins by constructing light cones from the remapped DMHC snapshots, which serve as the basis for simulating the 21 cm signal. The main goal is to compute the brightness temperature, which reflects the intensity of radiation from neutral hydrogen (HI). The light cone is built by stacking multiple snapshots from the halo catalog, accounting for the evolution of halo positions over time, redshift, or frequency. Using periodic boundary conditions to minimize issues such as duplication in large boxes and missing information in small ones,  deep light cone catalogs are generated by placing an observer at 100 random positions within the simulation box \citep{Zhang_2022}. For each redshift range, we filter the positions according to the corresponding comoving distances. Next, we calculate the ratio of the observer’s position relative to the comoving distance for the chosen redshift range and select halos based on their evolving positions, ensuring that halos from different redshifts are included. This method avoids relying on a single snapshot and provides a more accurate representation of the evolution of the halo. Each observer position yields a unique realization of the simulation, with snapshots stitched together to form light cone catalogs that capture cosmic evolution.

Each light cone catalog is then populated with neutral hydrogen (HI) by assigning HI mass to determine its distribution. At low redshift, as in our case, neutral hydrogen is mostly contained within galaxies and dark matter halos. In fact, the IllustrisTNG hydrodynamic simulation has shown that more than $95\%$ HI gas lies within the virial radius of dark matter halos \citep{Villaescusa_Navarro_2018}. 

Currently, there are several approaches to generating the HI distribution — some based on hydrodynamic simulations, and others using empirical HI mass–halo mass relations. \citep{Zhang_2022} discussed the advantages and disadvantages of these methods in terms of their accuracy and computational efficiency for different mock-building purposes. In this work, we adopt the HI halo occupation distribution (HOD) model from \citep{Zhang_2022}, which fits the HI mass–halo mass relation using the semi-analytical galaxy catalog of the ELUCID N-body simulation\citep{wang2016elucid, Luo2016}. The model accounts for central and satellite galaxies, combining a linear and Gaussian function with five free parameters: $a$, $b$, $c$, $d$, and $f$ \citep{Zhang_2022} 
 \begin{equation} \label{mhi}
    \log_{10}(\text{M}_{\text{HI}}) = a \log_{10}(\text{M}) + b+c e^{-\log_{10}(\text{M}) -d^2/f^2} \,,
\end{equation}
where the parameters are assumed to be linearly dependent on redshift in the range $0 < z < 0.66$:
\begin{equation}
 \begin{split}
 a &= 0.78 - 0.03\,z \nonumber \\
 b &= -0.23+0.68\,z \nonumber\\
 c &= 0.92 - 0.32\,z \nonumber\\
 d &= 11 + 0.038\,z \nonumber\\
 f &=0.79+0.18\,z \nonumber
 \end{split}
 \end{equation} 
with $\text{M}_{\text{HI}}$ and $\text{M}$ denoting the HI mass and halo mass, respectively, both in units of $\text{M}_{\odot} h^{-1}$. This relation provides a cosmology-independent way to populate the halos with HI, making it suitable for use across different cosmologies. Although more accurate fits could be obtained for each cosmology, the HOD method is sufficient for our purposes.

Once the HI gas mass is assigned to each halo, the HI density $\rho_{\text{HI}}$ is computed and related to the brightness temperature $\text{T}_\text{b}$ of emitted HI gas \citep{Villaescusa_Navarro_2018}
\begin{equation} \label{btemp}
    \text{T}_\text{b}=189 \, h \,\frac{H_0 (1+z)^2}{H(z)}\frac{\rho_{\text{HI}}}{\rho_c} \,,
\end{equation}
where $\rho_c$ is the critical density, $H(z)$ is the Hubble parameter, $H_0$ is the Hubble constant, and $\text{T}_\text{b}$ is in millikelvin (mK) units. To compute the volumetric HI density, we divide the halos into redshift bins (equally spaced in frequency) and apply HEALPix pixelization \citep{2005:Gorski} to the HI mass using \textit{nside} = 256. Finally, due to the possibility of overestimating the HI mass in halos compared to real observations (or underestimating it and missing some HI gas),it is necessary to rescale the maps using \citep{Zhang_2022}
   \begin{equation}
      \text{T}'_\text{b} = \text{T}_\text{b} \,\frac{\Omega'_{\text{HI}}}{\Omega_{\text{HI}}} \,.
 \end{equation}
Therefore, for a set of  halo catalogs at different redshifts, we obtain the brightness temperature of the 21cm signal. 

The methodology comprises the following steps: 
\begin{itemize}
    \item Construct Light Cones from the remapped DMHC snapshots. These light cones will serve as the foundation for simulating the 21 cm signal.
    \item Generate Deep Light Cone Catalogs: To minimize issues related to box size. This is done by placing an observer at 100 random positions within the simulation box and stacking multiple snapshots, accounting for the evolution of halo positions over time, redshift, or frequency. Periodic boundary conditions are utilized in this process.
    \item Populate Light Cone Catalogs with HI: Assign neutral hydrogen $(HI)$ mass to the halos within each light cone catalog to determine the distribution of HI applying the HI Halo Occupation Distribution (HOD) Model.
    \item Compute the HI Density once the HI gas mass is assigned to each halo.
    \item Calculate the Brightness Temperature using Equation~\eqref{btemp}.
    \item Rescale the Brightness Temperature Maps (Optional but Recommended): Rescale the brightness temperature maps using Equation \eqref{btemp} with the ratio of $\Omega_{HI}'$ to $\Omega_{HI}$ to account for potential overestimation or underestimation of HI mass in halos compared to observations.
    \item Obtain Brightness Temperature Maps: After these steps, for a given set of halo catalogs at different redshifts, the brightness temperature of the 21cm signal is obtained.
\end{itemize}

%%%%%%%%%%%%%%%%%%%%%%%%%%%%%%%%%%%%%%%%
\section{Simulations}\label{sec:data}
%%%%%%%%%%%%%%%%%%%%%%%%%%%%%%%%%%%%%%%%
%Remapping offers an efficient alternative to running new N-body simulations. 

In this work, we focus on applying remapping to the \emph{Horizon Run 4} (HR4) \citep{hr4} and consider it as the original N-body simulation, which is remapped to other cosmologies.In order to compar the remapped outputs, we use the \emph{OuterRim} and  \emph{Miratitan}  simulations. This allows us to demonstrate how remapping can reproduce the target cosmologies. Table \ref{tab:simulations} lists the simulations used for implementation and testing, which are discussed in the following sections. 

\subsection{Horizon Run 4 simulation}

The \emph{Horizon Run 4} (HR4)  \citep{hr4} was chosen as the original N-body simulation due to its substantial box size and availability at various redshifts. Its large volume allows the generation of 21 cm mock data over a wide frequency range that matches the redshift interval targeted by the BINGO telescope team \citep{Abdalla2022}, $0.127 < z < 0.449$ (and $980 - 1260\,MHz$). In fact, it has been shown that when the simulation box exceeds $2400\,h^{-1}\mathrm{Mpc}$, a full-sky light cone within this redshift interval can be constructed without relying on periodic boundary conditions \citep{Zhang_2022}.

The HR4 catalogs are the result of cosmological N-body simulations adopting the WMAP5 cosmology \citep{Dunkley2009}, containing $6300^3$ particles with a minimum mass of $\text{M}_{\text{min}} = 2.7 \times 10^{11} h^{-1}\text{M}_{\odot}$ in a cubic box of comoving length $L_{\text{HR4}} = 3150 h^{-1}\,\mathrm{Mpc}$, spanning the redshift range $0<z<0.5$, divided into $7$ bins \citep{hr4}. It utilized an improved version of the GOTPM code \citep{dubinski2004gotpm}, which incorporated the CAMB package \citep{lewis2011camb} to compute the input power spectrum. The initial positions and velocities of the particles were determined using the second-order Lagrangian perturbation theory (2LPT) method proposed in \cite{jenkins2010second}. 

The gravitational force was computed by splitting the Newtonian force law into long- and short-range components, with the long-range forces computed from the Poisson equation in Fourier space using the Particle-Mesh (PM) method, and the short-range forces computed with the Tree method. The code was parallelized using MPI~\footnote{\url{https://www.mpi-forum.org/docs/}} and OpenMP~\footnote{\url{https://www.openmp.org/specifications/}} with a one-dimensional domain decomposition in the slabs along the z-direction, adopting a dynamic domain decomposition approach to ensure equal particle count within each domain. However, slab domains have larger surface-to-volume ratios compared to cubic domains, which results in larger communication sizes between domains.

\begin{table}
\caption{Table for Original and Target Cosmology and Simulation Names used for the validation comparison.}
\footnotesize
\centering
\begin{tabular}{|l|c|c|c|}
\hline
\multicolumn{2}{|c|}{Original}  & \multicolumn{2}{|c|}{Target} \\
Cosmology & Simulation    & Cosmology & Simulation \\ \hline
\multirow{2}{*}{WMAP5} & \multirow{2}{*}{HR4} & WMAP7    & OuterRim \\
                       &                       & CPL      & MiraTitan (M003) \\ \hline
\end{tabular}
\label{tab:simulations}
\end{table}

% \begin{table}
% \centering
% \begin{tabular}{|l|c|c|c|}
% \hline
% \multicolumn{2}{|c|}{Original}  & \multicolumn{2}{|c|}{Target} \\
% Cosmology & Simulation    & Cosmology & Simulation \\ \hline
% \multirow{WMAP5} & \multirow{HR4} & WMAP7    & OuterRim \\
%                       % &  & Planck18 & NewWorlds (QonoS) \\ 
%                       &  & CPL      & MiraTitan (M003) \\ \hline
% \end{tabular}
% \caption{Table for Original and Target Cosmology and Simulation Names used for the validation comparison.}
% \label{tab:simulations}
% \end{table}
%
\subsection{OuterRim and Miratitan  simulations}

The choice of target cosmologies for this study was primarily driven by two key factors: the availability of N-body simulations with comparable numerical precision to our remapped catalogs, and the availability of these simulations at the target redshift range relevant for our comparisons. Consequently, we focus on the WMAP7 and CPL cosmologies, utilizing specific high-resolution simulations. The remapped results for the WMAP7 cosmology are contrasted with the \emph{OuterRim} simulation, while the remapped results for the CPL cosmology are compared against the \emph{Mira-Titan} (M003) simulation \citep{Heitmann_2016}. In the case of Planck2018, we did not find a simulation satisfying both factors above. Therefore, the remapped results are compared with the theoretical power spectrum, a standard practice for validating cosmological models.

The \emph{OuterRim} simulation\footnote{\url{https://cosmology.alcf.anl.gov/outerrim}} \citep{Heitmann_2019} was carried out on the Mira supercomputer at the Argonne Leadership Computing Facility. It covers  $(4.225 \, h^{-1}\mathrm{Gpc})^3$, a substantial comoving volume, and evolves $10240^3$ dark matter particles. The cosmological parameters adopted are close to the best-fit WMAP7 cosmology \citep{Komatsu2011}. This large volume and high resolution make it one of the most extensive cosmological simulations available. Outputs from the \emph{OuterRim} simulation, generated using the HACC (Hardware/Hybrid Accelerated Cosmology Code), have been utilized in various scientific investigations.

The \emph{Mira-Titan Universe} simulations\footnote{\url{https://cosmology.alcf.anl.gov/transfer/miratitan}} were  executed on the Mira and Titan supercomputers at the Argonne Leadership Computing Facility and the Oak Ridge National Laboratory. These simulations encompass a variety of cosmological models, including those featuring a dynamical dark energy equation of state parameterized by

\begin{equation}
w(z) = \omega_0 + \omega_a \frac{z}{1+z} \,.
\end{equation}

Each simulation within the suite covers a comoving volume of $(2.1 \, h^{-1}\mathrm{Gpc})^3$ and evolves $3200^3$ dark matter particles. Outputs, including halo catalogs, are available at 27 distinct redshifts spanning the range $0 \le z \le 4$, ensuring coverage of our target redshift range. For the CPL cosmology, we specifically utilized the M003 simulation from this group of simulations, which implements a CPL dark energy model \citep{Chevallier2001,Linder2003} with specific values for $\omega_0$ and $\omega_a$ as defined in its setup.

The direct comparison between our WMAP7 remapped halo catalogs and the halo catalogs from the \emph{OuterRim} simulation, and between our CPL remapped catalogs and the halo catalogs from the \emph{Mira-Titan M003} simulation, both within the relevant redshift range, allows us to assess the accuracy and reliability of the remapping technique in reproducing key large-scale structure statistics across different cosmological scenarios. The large volumes of these simulations also minimize the impact of cosmic variance on our comparisons.

%%%%%%%%%%%%%%%%%%%%%%%%%%%%%%%%%%%%%
\section{Results}\label{sec:results}
%%%%%%%%%%%%%%%%%%%%%%%%%%%%%%%%%%%%%
\subsection{Remapping Dark Matter Halo Catalogs}
%%%%%%%%%%%%%%%%%%%%%%%%%%%%%%%%%%%%%
First, we applied the remapping method described above to remap the WMAP5 cosmology to WMAP7, CPL and Planck2018 cosmologies defined by the parameters shown in the Table~\ref{tab:param}. The minimization process using Eq.~(\ref{rms}) was performed and the results are summarized in Table~\ref{tab:minimization_param}, which list also  the rescaling parameters $s$ (distance and position), $s_m$ (mass) for each target cosmology, as well as the scales L and L' for the box lenghts.  

\begin{table*} %[h!]
 \caption {The cosmological parameters for each cosmological model adopted in this work to demonstrate examples that we remap to.} \label{tab:param} 
      \footnotesize

   \begin{center}
\begin{tabular}{|l|l|l|l|l|l|l|l|l|l|}
        \hline 
         $Cosmology/Model$ & $h$ &$\Omega_m$ & $\Omega_b$  &$\sigma_8$ &$n_s$ & $\omega_0$ & $\omega_a$ \\
         \hline \hline
         WMAP5 & $0.72$ & $0.26$ & $0.043$ & $0.793$ & $0.96$ & $-1.0$  & $0$  \\
         \hline
         WMAP7 & $0.71$ & $0.266$ & $0.044$ & $0.8$ & $0.963$ & $-1.0$ & $0$ \\
         \hline
         Planck18 & $0.676$ & $0.309$ & $0.0489$ & $0.8102 $ & $0.9665$ & $-1.0$ & $0$ \\
         \hline
         CPL(M003)& $0.7167$ & $0.3017$ & $0.0427$ & $0.9$ & $0.8944$ & $-1.10$ & $-0.2833$ \\
         \hline
  \end {tabular}
   
 \end{center}
\end {table*}

\begin{table}
\caption{The final results of the remapping process showing for each cosmological model we adopted for remapping starting from the wmap5 HR4 catalogs as original, with the original redshift $z$ we used and the target redshift we were aiming for $z'$ for each alongside with the rescaling parameters $s$ for distances and positions, and the parameter $s_m$ for mass rescaling.
}
\label{tab:minimization_param}
     \footnotesize

  \centering
    \begin{tabular}{|l|l|l|l|l|l|l|l}
        \hline 
        Cosmology & $z'$ & $z$ & $s$ & $s_m$ & $L$ & $L'$ \\
         \hline 
         $\text{WMAP 7}$ & $0.43$ & $0.4$ & $0.9873$ & $0.9845$ & 3150 & 3109,9 \\
         $\text{Planck2018}$ &  $0.502$ & $0.291$ & $0.8616$ &  $0.7601$ & 3150 & 2714,1\\
         $\text{CPL}$  &  $0.401$ & $0.153$ & $0.9874$ & $1.117$ & 3150 & 3110.3\\
         \hline
    \end{tabular}     
\end{table}

\begin{figure}[ht]
    \centering
    
    \includegraphics[height=7cm]{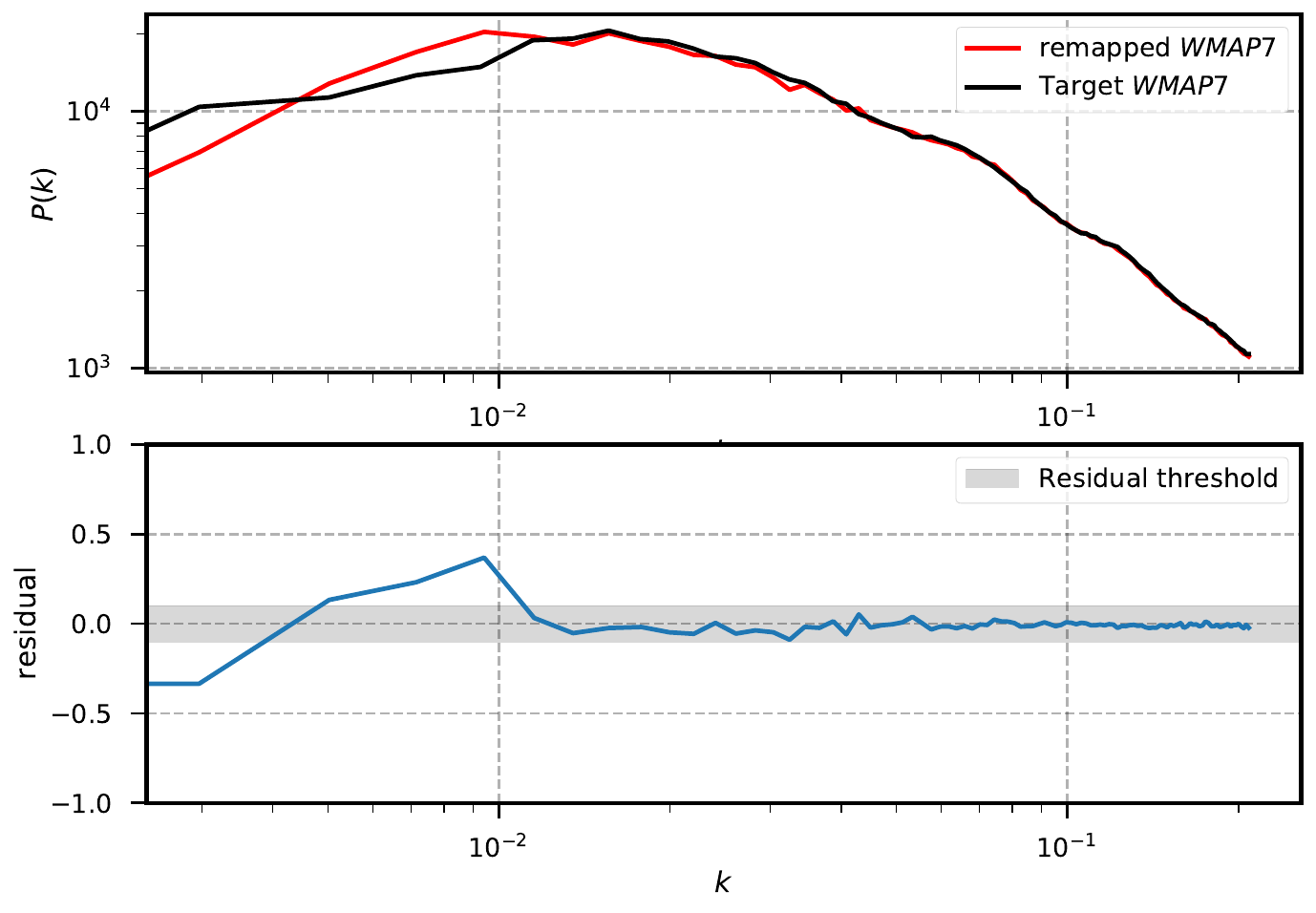}
    \caption{Top: $P(k)$ comparison between remapped DMHC of wmap5 into wmap7, together with $P(k)$ obtained from DMHC of the wmap7 cosmological model represented in the OuterRim simulations. 
    Bottom: the residuals between the remapped and WMAP7 OuterRim power spectra show excellent consistency, not exceeding $5\%$ for scales $k \gtrsim 10^{-2}$h/Mpc}
    \label{fig:pk_wmap7_lim}
\end{figure} 

\begin{figure}[ht]
    \centering

    \includegraphics[height=7cm]{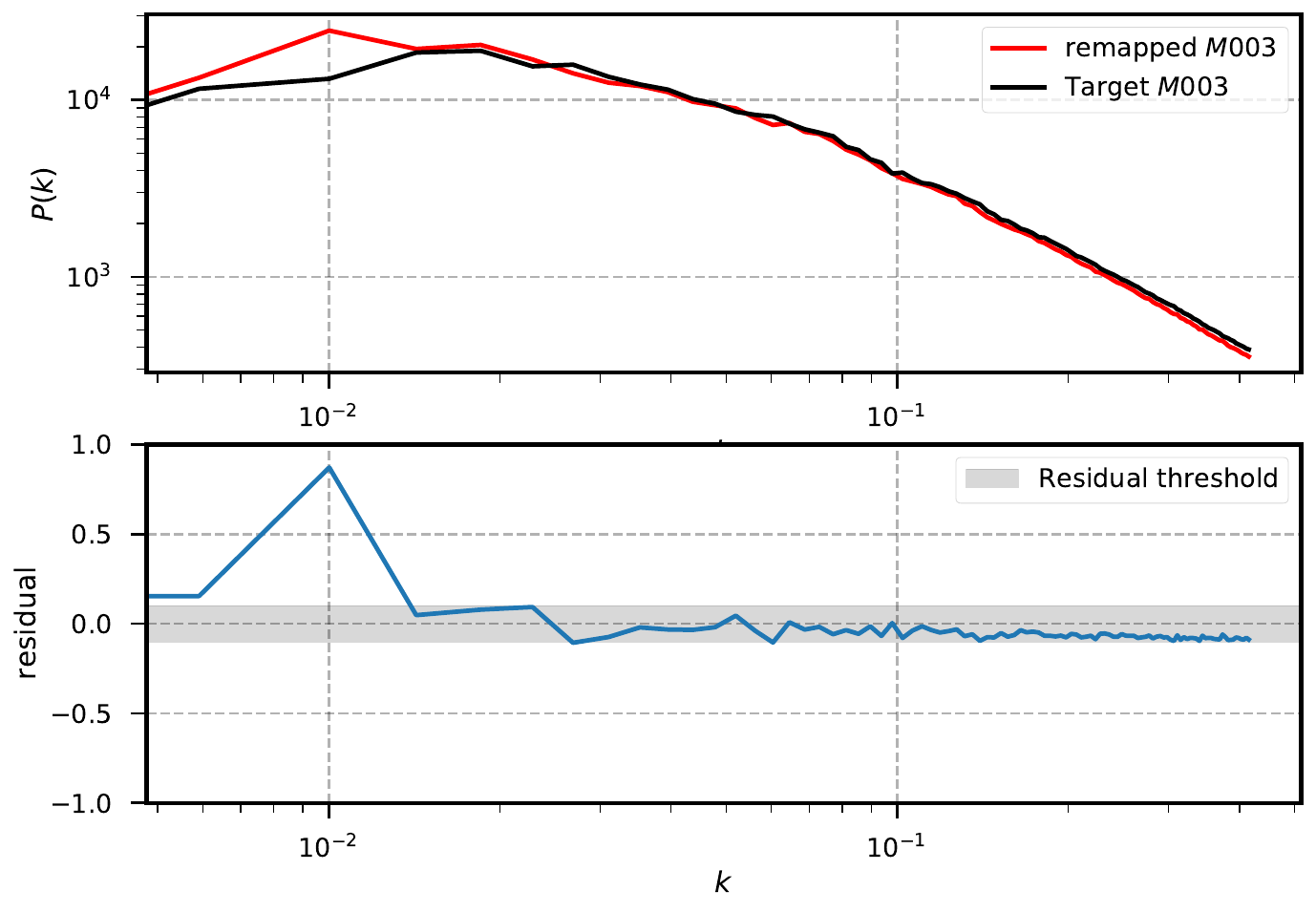}
    
    \caption{Top: P(k) comparison between remapped simulation from WMAP5 into CPL, with the P(k) obtained from a simulation representing the CPL parametrization simulation M003 obtained from the MiraTitan simulations.
    Bottom: the residuals between the remapped WMAP5 and the MiraTitan CPL simulations also not  exceed $\sim \pm 5\%$ for scales $k \gtrsim 10^{-2}$h/Mpc, validating the remapping approach.
    }
    \label{fig:m003_sim}
\end{figure}

\begin{figure}[htpb]
\centering
\begin{minipage}[t]{\textwidth}
    \includegraphics[width=0.32\linewidth]{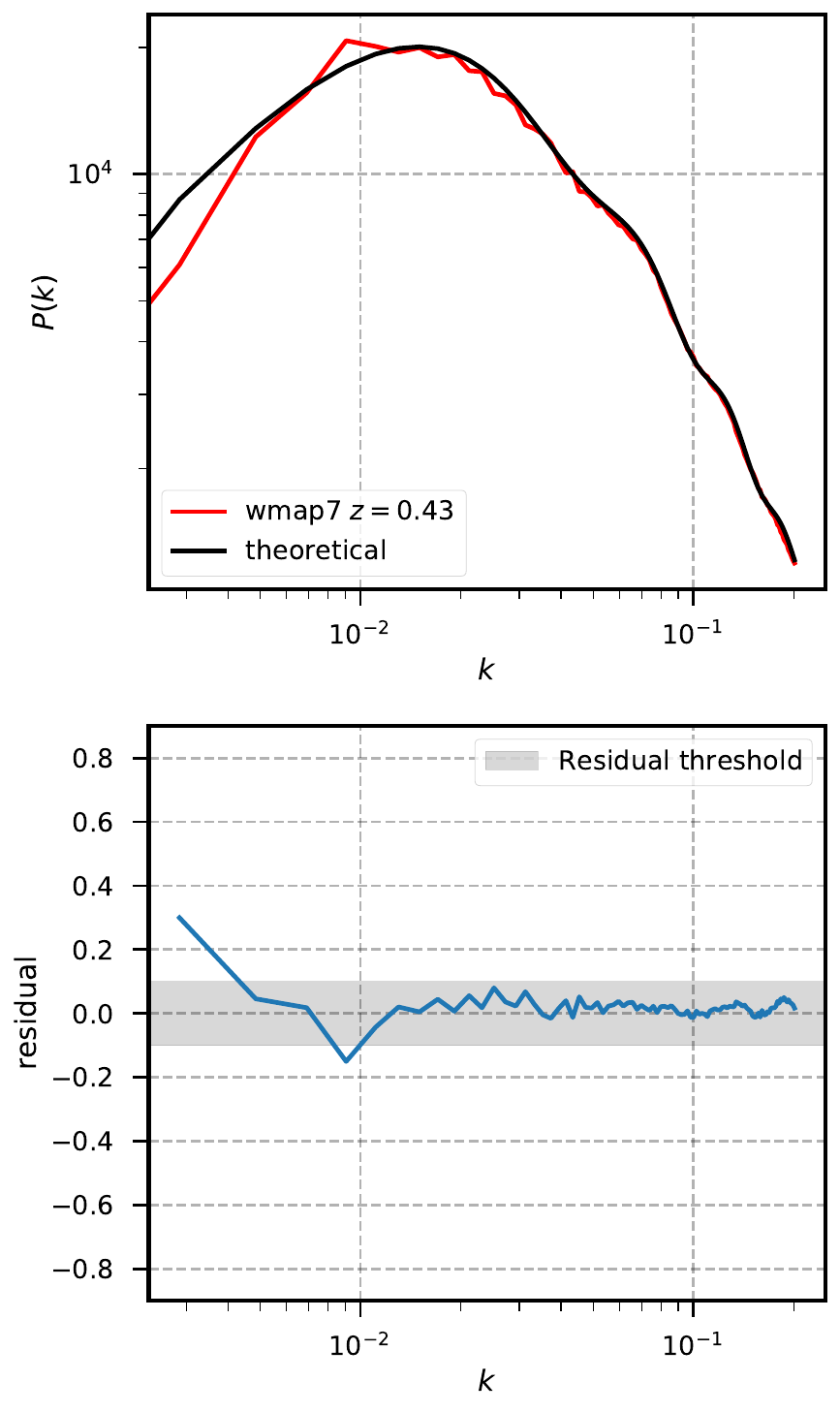}
  \includegraphics[width=0.32\linewidth]{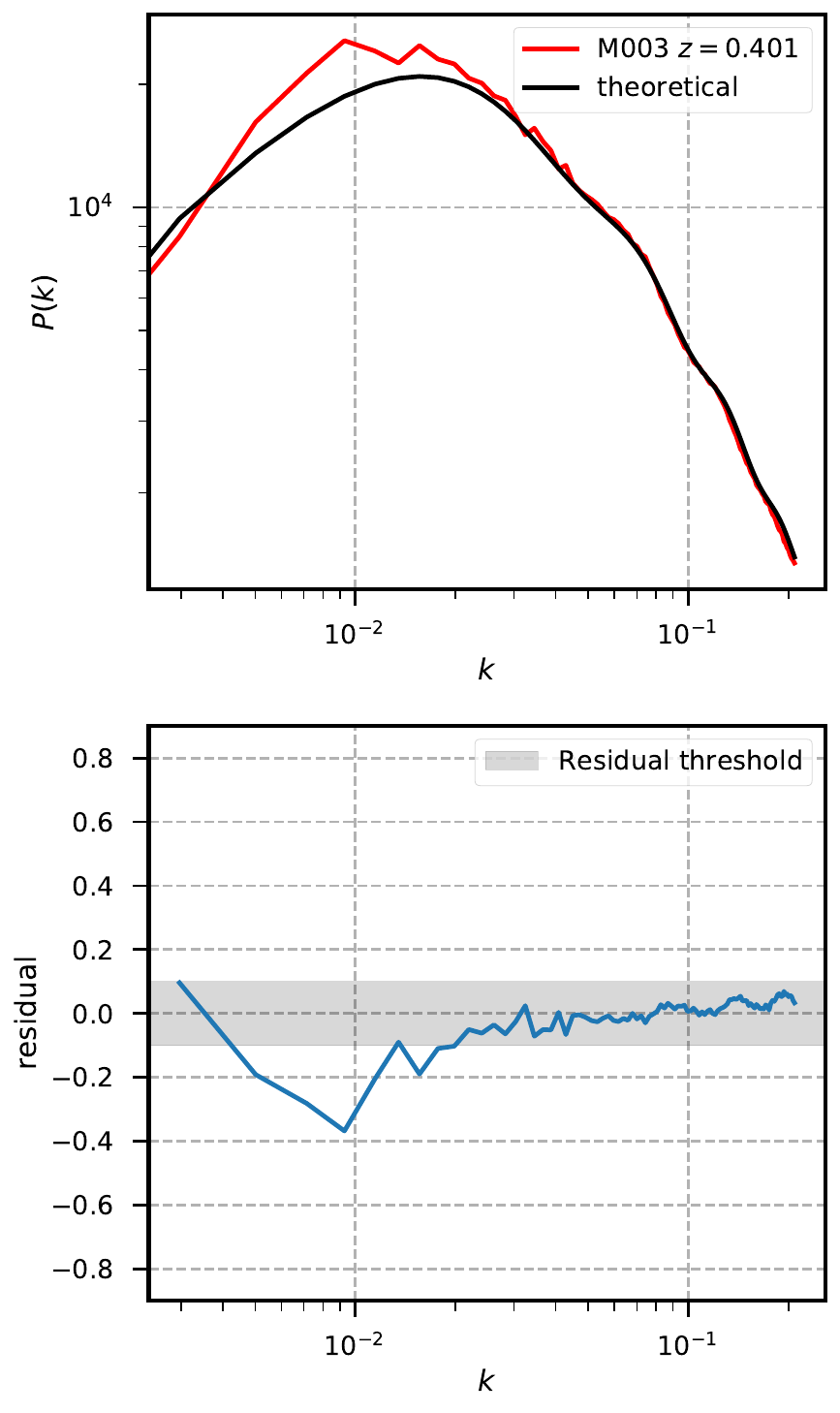}
      \includegraphics[width=0.32\linewidth]{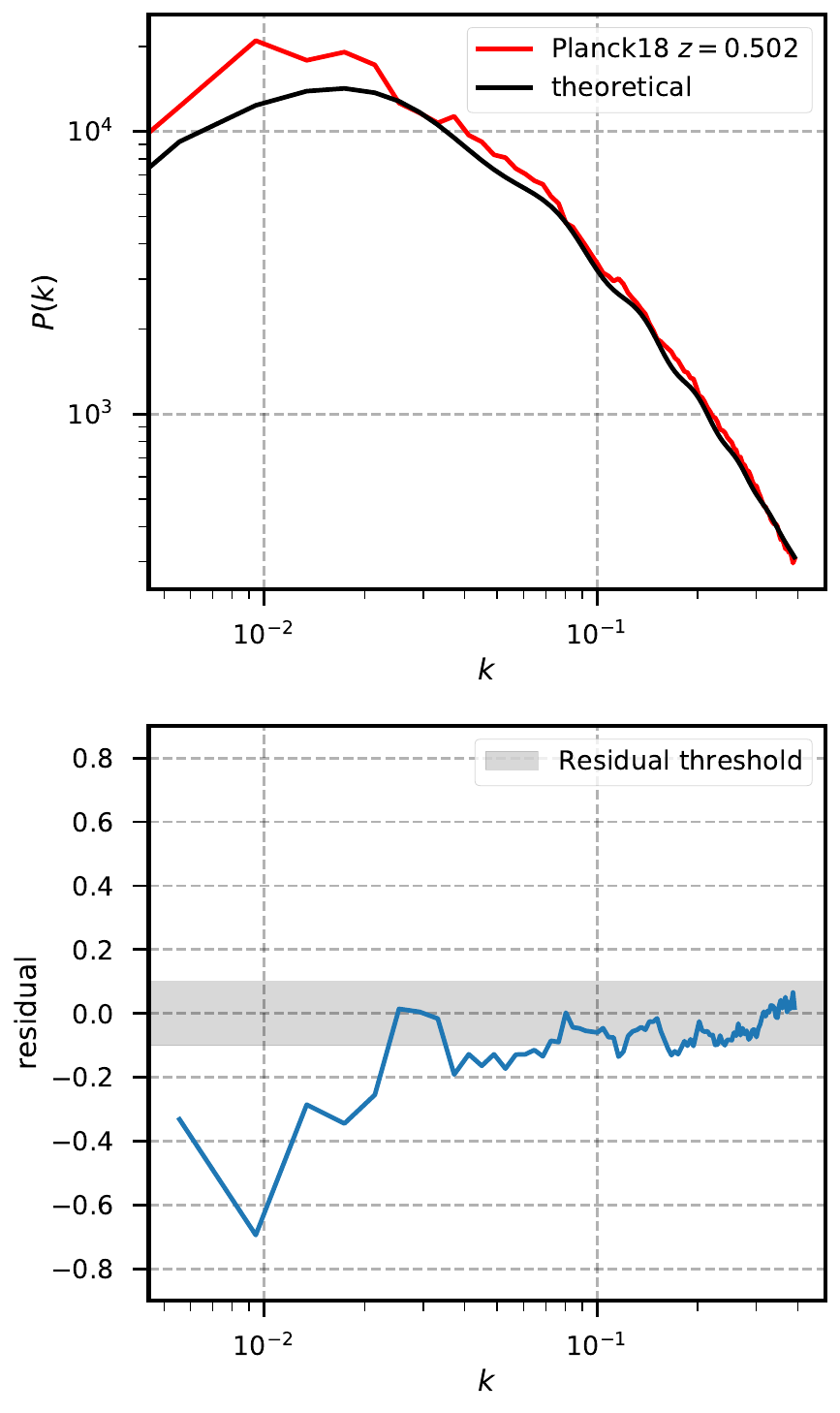}

\end{minipage}%

\caption{Matter power spectra $P(k)$ from remapped halo catalogs (red lines) for three cosmological models (from left to right: WMAP7, CPL, Planck18) compared to the corresponding theoretical power spectra (black lines). Lower panels show the fractional residuals $[P(k)_{\text{remapped}} - P(k)_{\text{theoretical}}] / P(k)_{\text{theoretical}}$ (blue lines).  This figure illustrates how the remapping successfully preserves the matter power spectrum in the linear regime, as indicated by the small fractional residuals shown in the lower panels.}\label{fig:theo_comp_cosmologies}

\end{figure}

We then shift the positions  using Eq.(\ref{shift}) to reproduce the target cosmologies  and generate the correponding dark matter halo catalogs.  These remapped catalogs are  compared with their counterparts from the \emph{OuterRim} and \emph{Mira Titan Universe} simulations. The results, shown in  Figs. \ref{fig:pk_wmap7_lim} and~\ref{fig:m003_sim}, contrast the halo  power spectra from the remapped and simulations catalogs. As shown in the figures, the remapped WMAP7 and CPL power spectra closely match their N-body counterparts, with residuals generally within $5\%$ across a broad range of scales. Deviations are slightly larger at larger scales (lower wavenumbers $<10^{-2}$), likely due to cosmic variance, and also increase at small scales due to nonlinear effects not fully captured by the remapping methodology. These results highlight the effectiveness of the remapping technique in reproducing the large-scale clustering features of the target cosmologies. The limitations of the method are discussed in detail in Appendix~\ref{Remapping precision}. To improve the results in the nonlinear regime, an additional step (halo reconstitution, discussed in \citep{Mead2014a}) is required. However, since our focus is on linear scales, this step is not applied here.

In addition to halo power spectrum comparisons, we present the matter power spectra $P(k)$  derived from the debiased remapped halo spectra and compare them with the theoretical predictions for each target cosmology . This allows us to assess the accuracy of the remapping in reproducing the underlying matter distribution. As shown in Fig.~\ref{fig:theo_comp_cosmologies}, this time including Planck2018 cosmology,  the residuals between the simulated matter power spectra and the theoretical predictions remain consistently below $10\%$. This accuracy achieved across all three cosmologies is sufficient for our purposes and  supports the reliability  of the remapping method implemented here in reproducing the large-scale structure of different cosmological models. Therefore, the remapping methodology successfully transformed the DMHC from the WMAP5 cosmology into three target cosmologies: WMAP7, Planck2018, and CPL. These remapped catalogs were used to generate 21cm brightness temperature maps.

%%%%%%%%%%%%%%%%%%%%%%%%%%%%%%%%%%%%%%%%%%%%%%5555
\subsection{Brightness Temperature Maps}
%%%%%%%%%%%%%%%%%%%%%%%%%%%%%%%%%%%%%%%%%%%%%%%

We applied the method to generate synthetic 21cm intensity maps, following the halo occupation distribution (HOD) approach described in Section~\ref{21cm}. Accordingly, we first constructed the light-cone catalogs using the previously remapped DMHC (WMAP7, Planck2018, and CPL) and populated them with neutral hydrogen (HI) mass to create brightness temperature maps, as described previously. As a benchmark, we performed the same HOD-based 21cm mock generation, and the same procedure was applied to \emph{OuterRim} and \emph{Mira Titan Universe} DMHC for comparison. These maps illustrate the spatial distribution of the 21cm signal.

\begin{figure*} %%[htbp]
\centering
\begin{minipage}{0.35\linewidth}
\centering
\includegraphics[width=\linewidth]{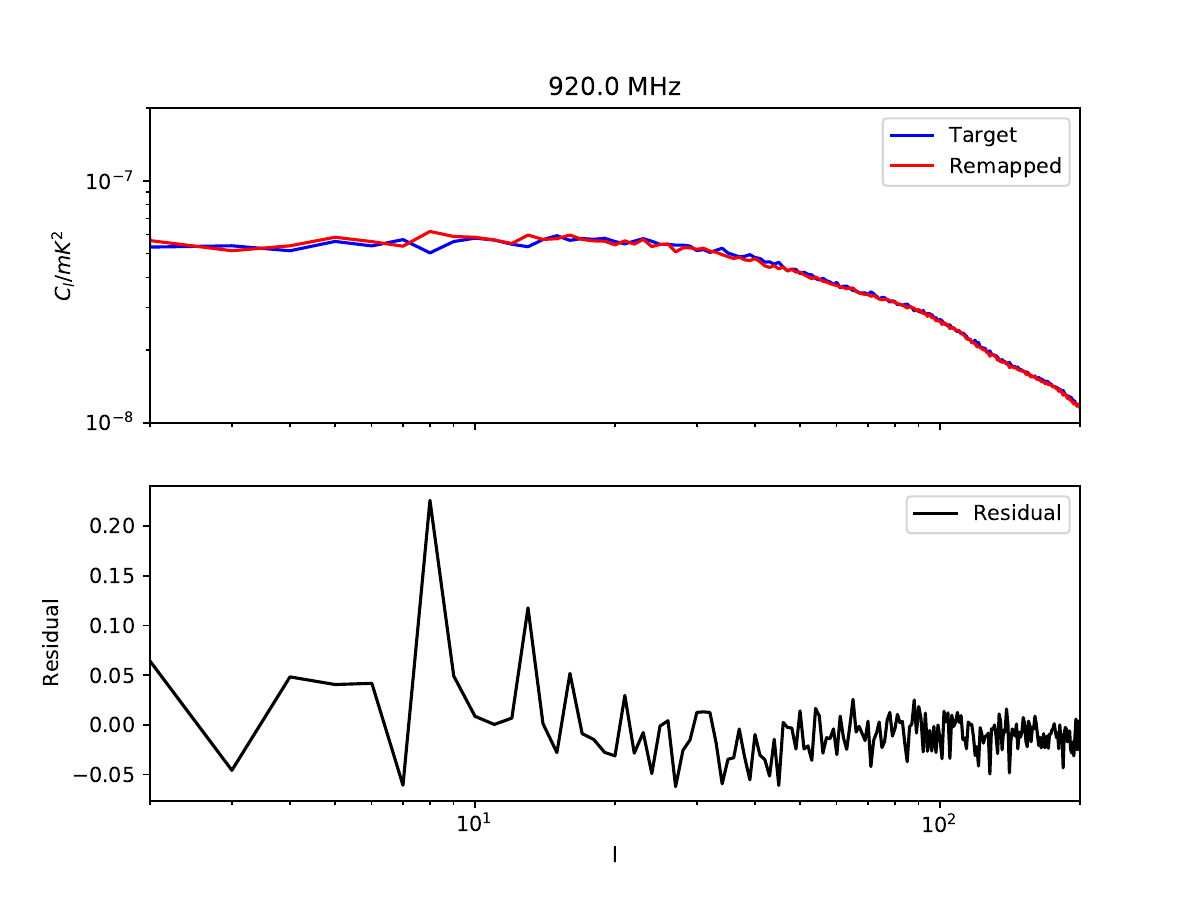}
\includegraphics[width=\linewidth]{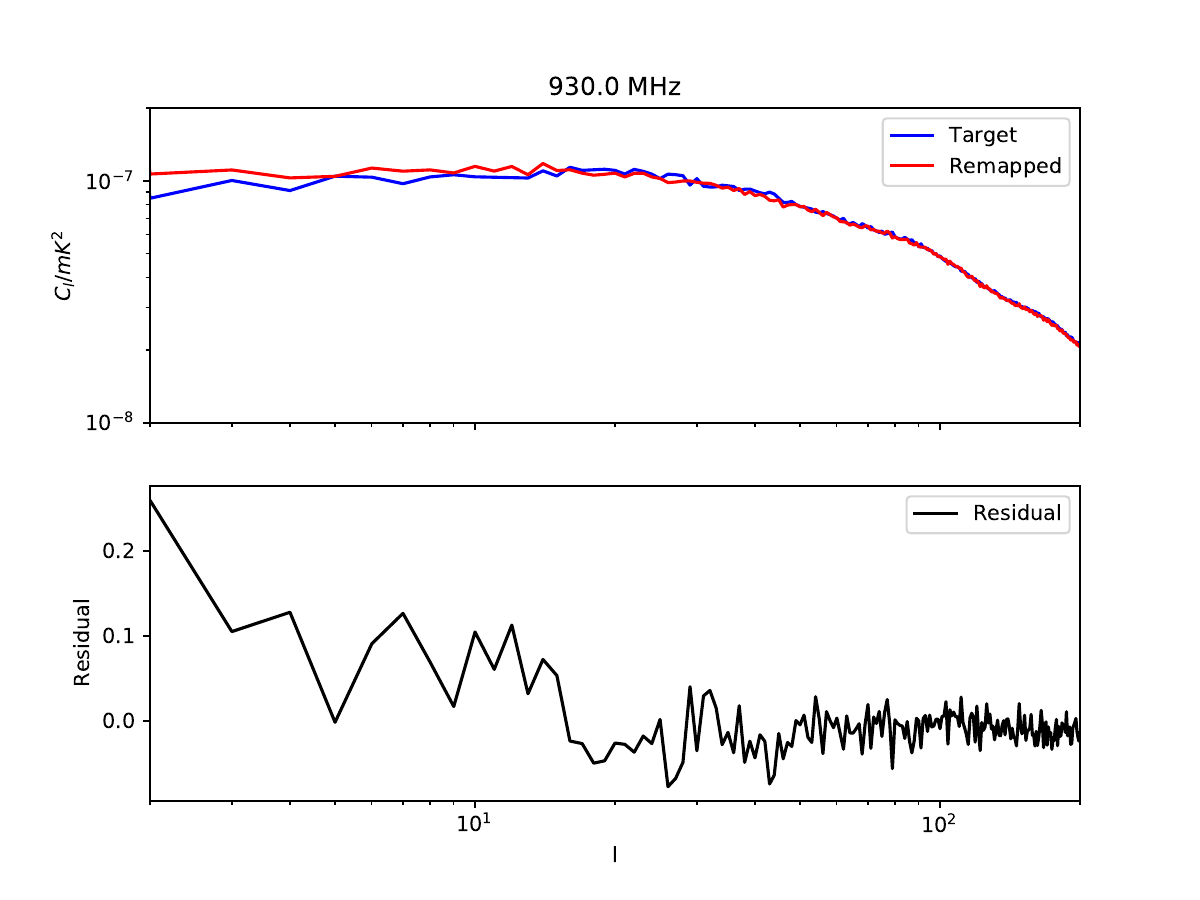}
\end{minipage}%
\begin{minipage}{0.35\linewidth}
\centering
\includegraphics[width=\linewidth]{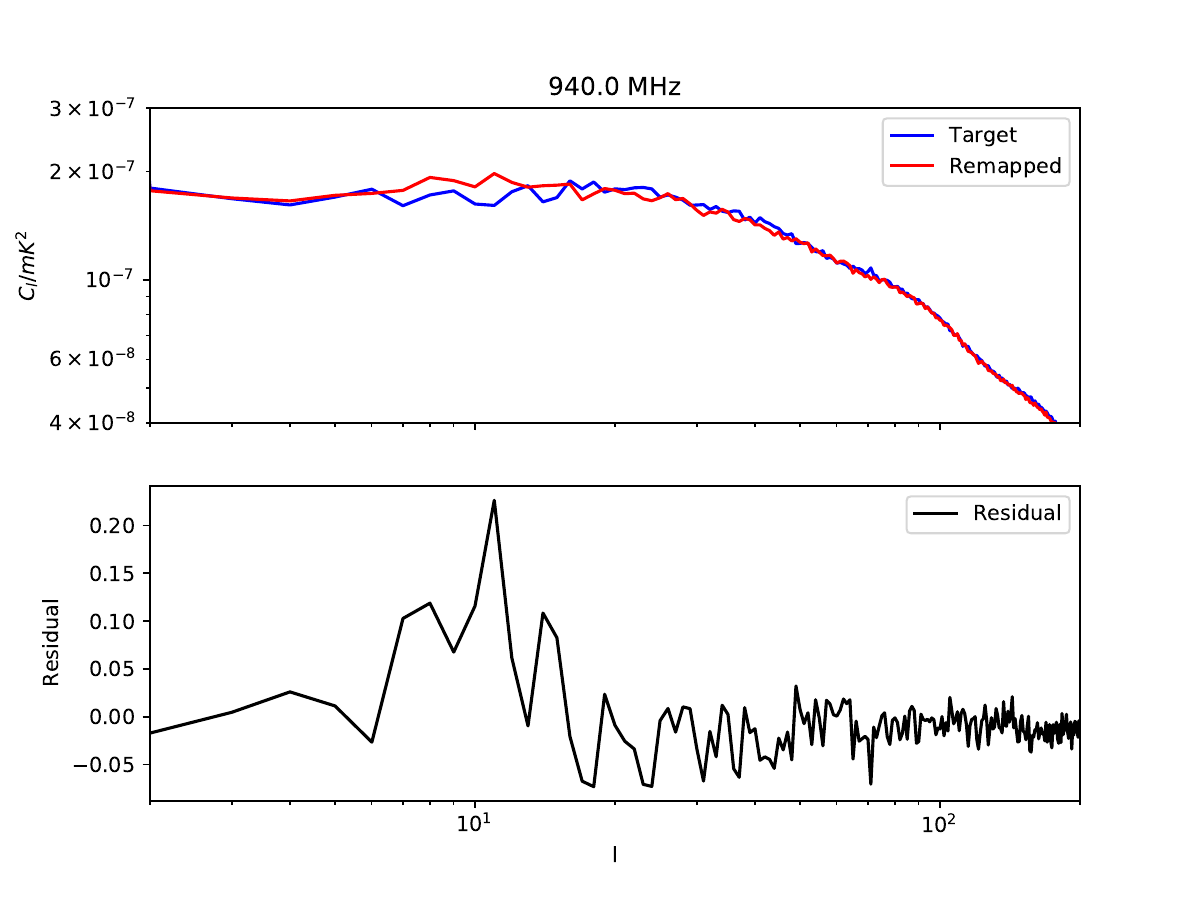}
\includegraphics[width=\linewidth]{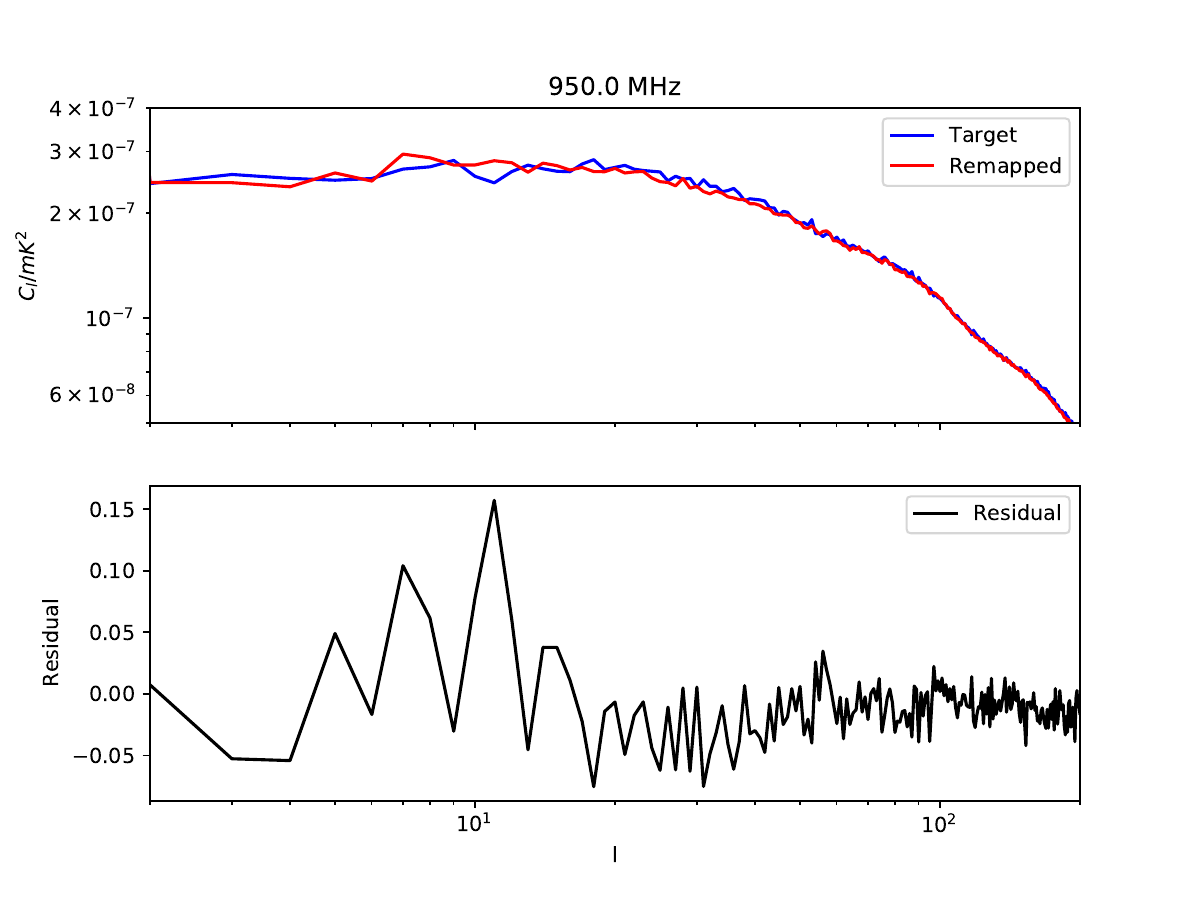}
\end{minipage}%
\begin{minipage}{0.35\linewidth}
\centering
\includegraphics[width=\linewidth]{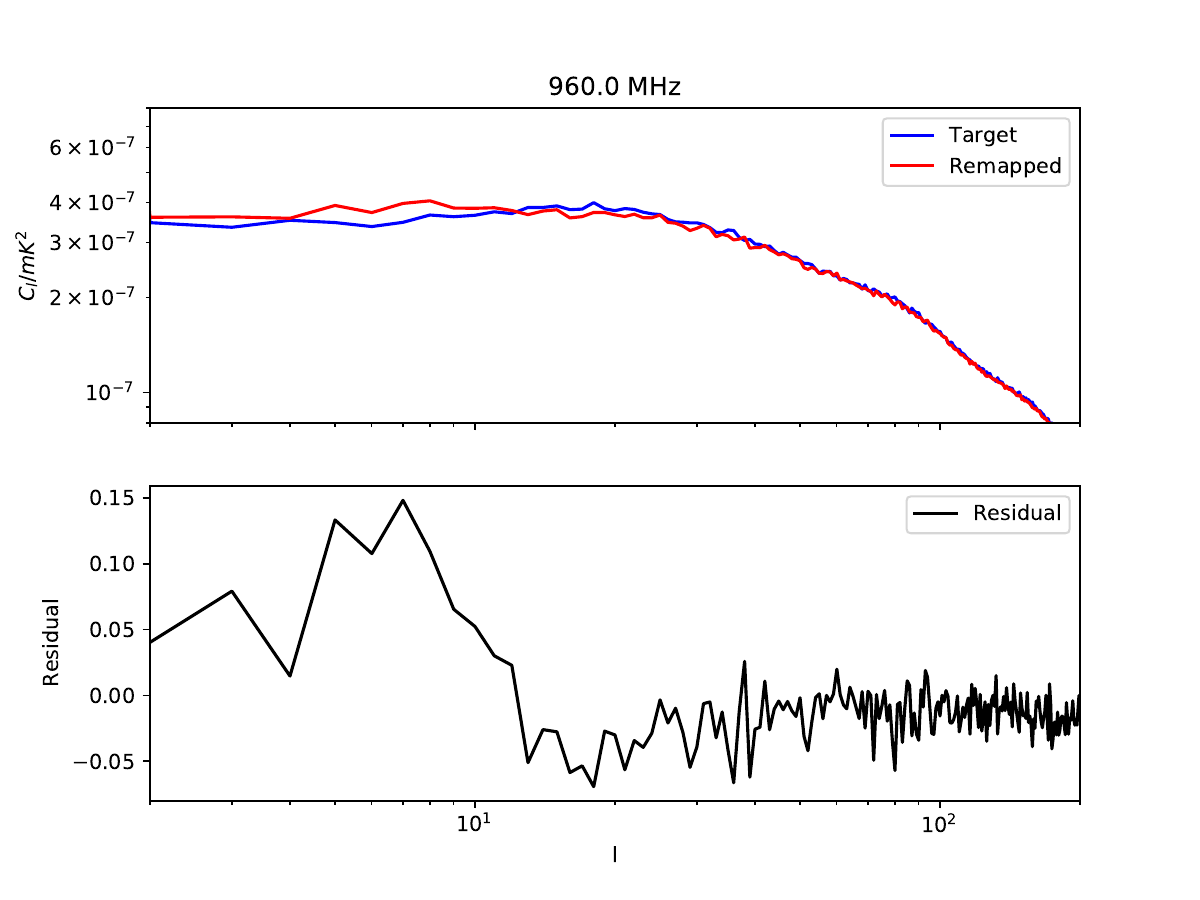}
\includegraphics[width=\linewidth]{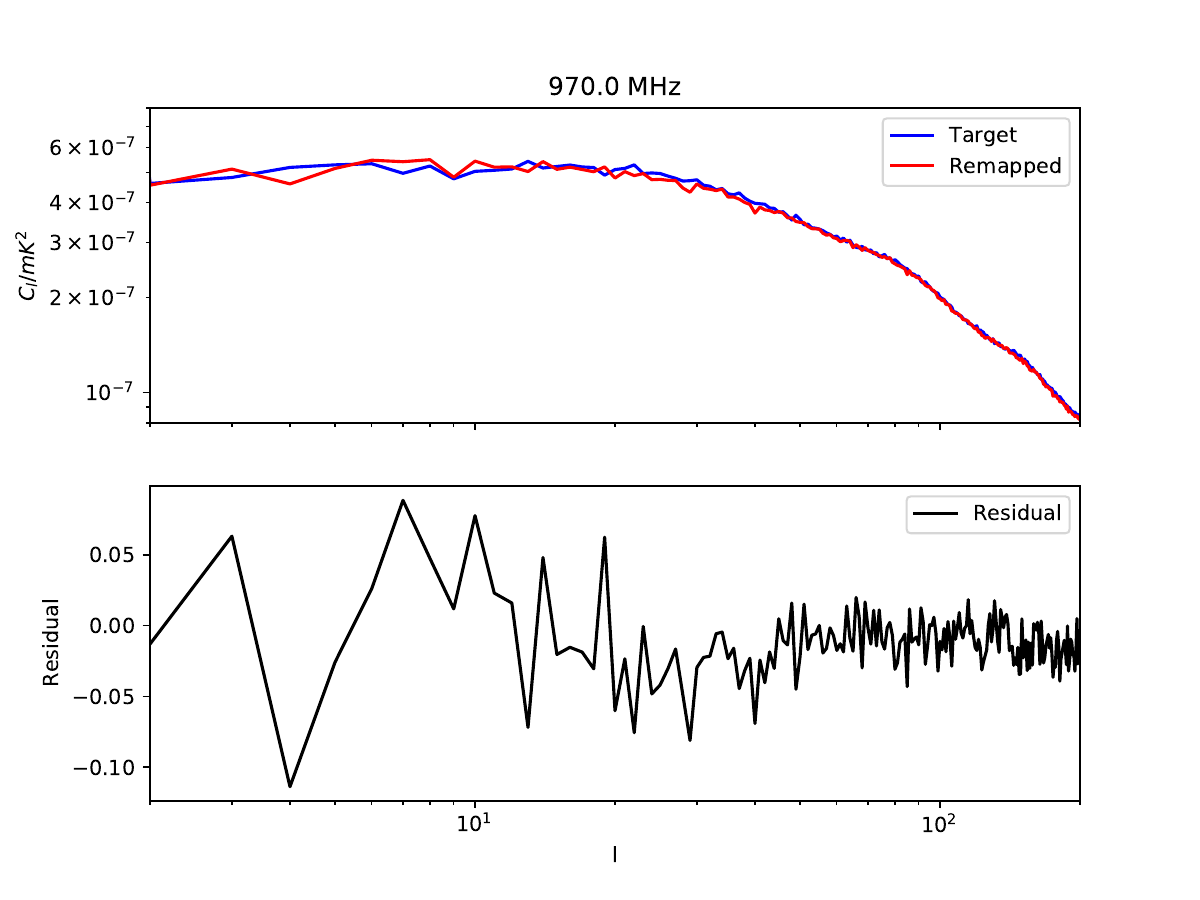}
\end{minipage}
\caption{Mean 21 cm $C_{\ell}$ obtained from the mocks generated using the remapped DMHC from the WMAP5 into WMAP7 cosmology (in red), compared to the $C_{\ell}$ from the mocks generated using the DMHC from the OuterRim simulations respresenting the WMAP7 cosmology (in blue). The key takeaway here is the minimal residual, which stays within $5\%$ to $10\%$, proving that the remapping precision is maintained even at the level of 21 cm $C_{\ell}$ from the mocks. This strong agreement highlights the reliability of using remapping for future 21 cm mock simulations.}
\label{fig:cl_wmap7_mean}
\end{figure*}

\begin{figure*} %%[htbp]
\centering
\begin{minipage}{0.52\linewidth}
\centering
\includegraphics[width=\linewidth]{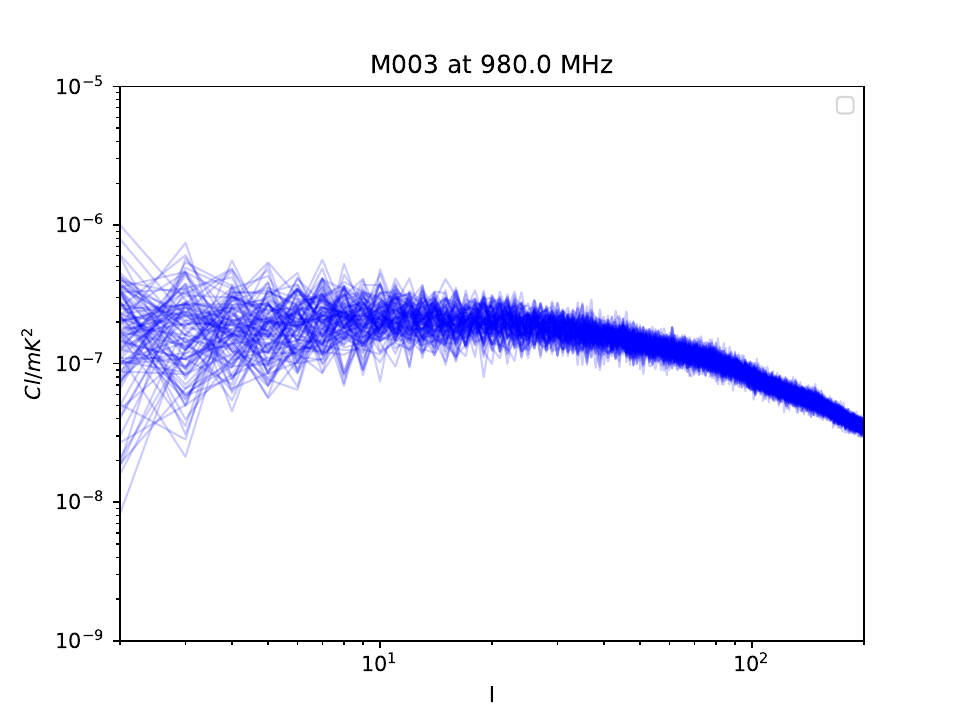}
\includegraphics[width=\linewidth]{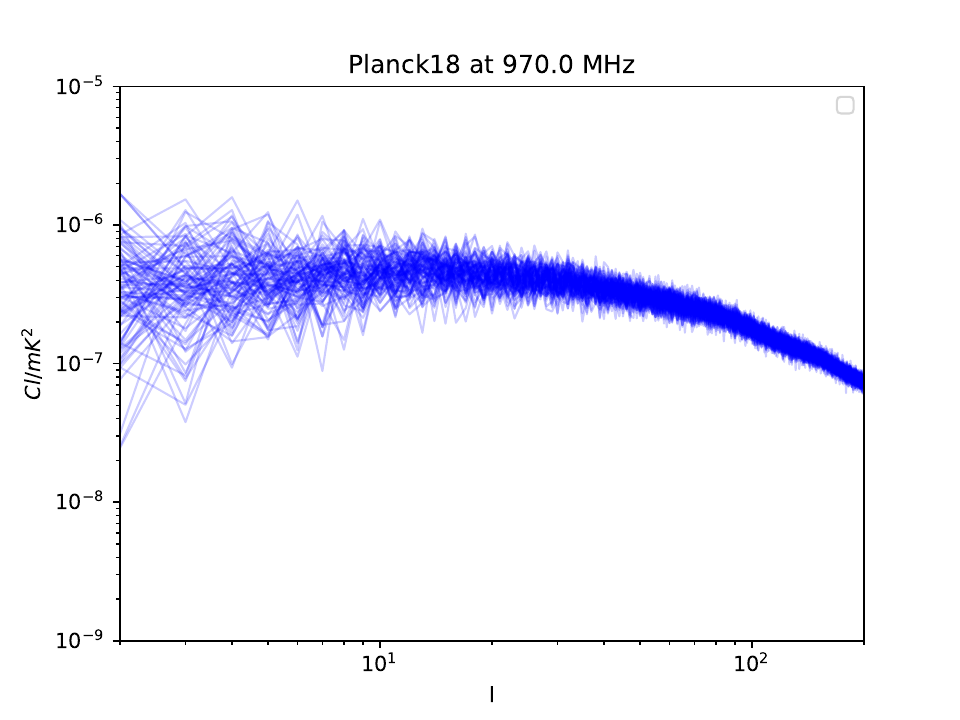}
\end{minipage}%
\begin{minipage}{0.52\linewidth}
\centering
\includegraphics[width=\linewidth]{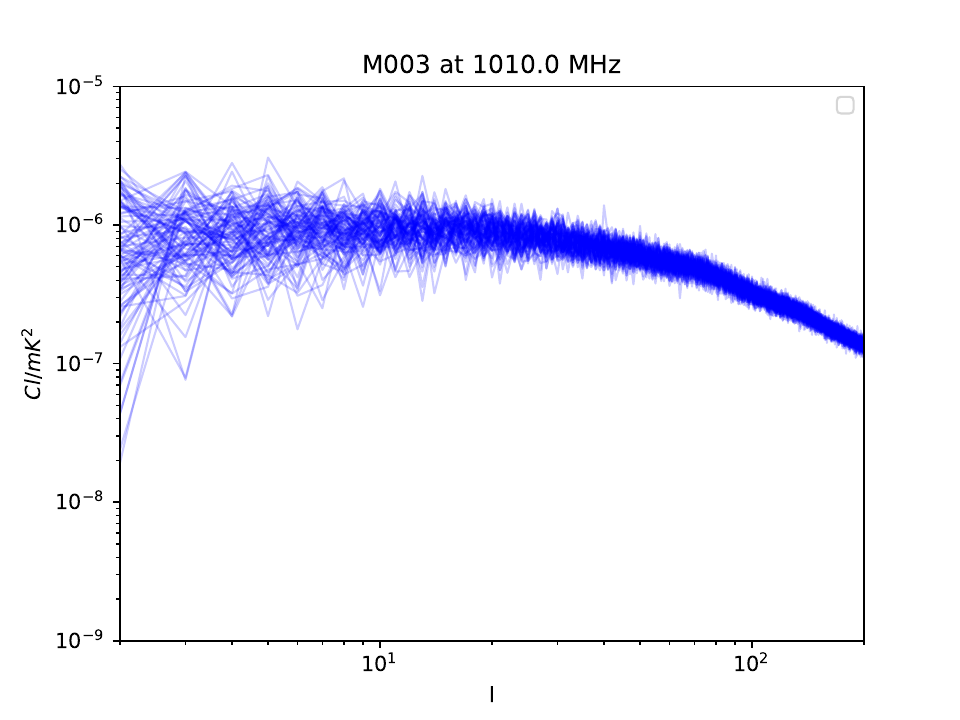}
\includegraphics[width=\linewidth]{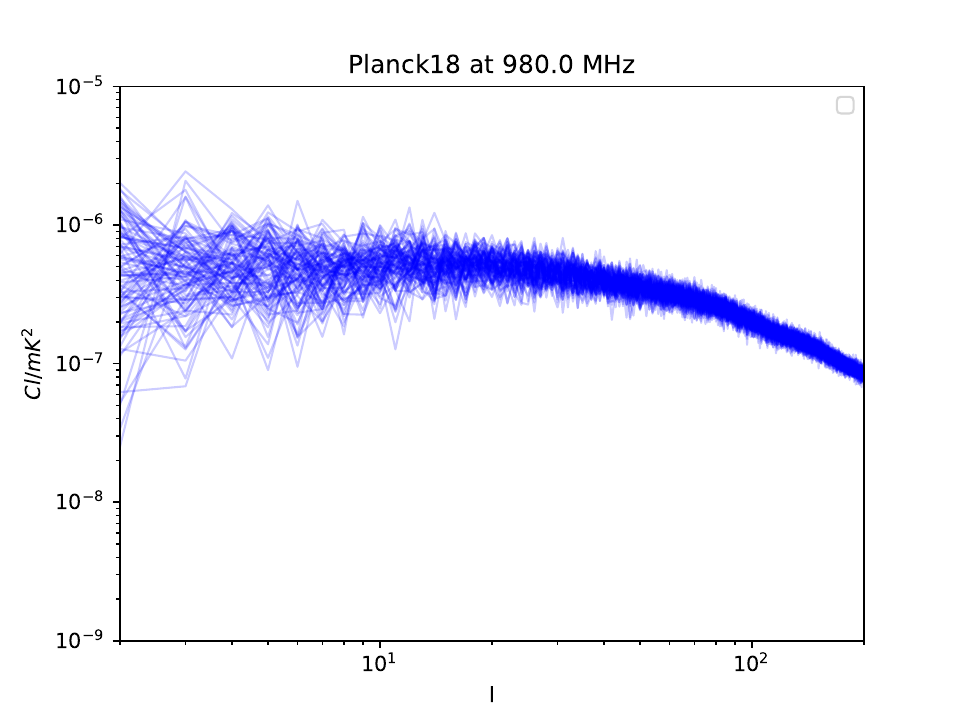}
\end{minipage}%
\caption{21 cm $C_{\ell}$ obtained from the mocks generated using the remapped DMHCs from the wmap5 into the M003 cosmology, on the upper line, and into Planck18 in the lower line, frequencies are mentioned in the titles. }
\label{figure:cls_m003}
\end{figure*}

To quantitatively assess the effectiveness of the remapping technique in the context of  21cm intensity mapping, we focused on comparison  between the angular power spectra ($C_{\ell}$) of these maps.  To mitigate the impact of cosmic variance, we generated 100 independent realizations of 21cm mocks for both the remapped WMAP7 catalog and the \emph{OuterRim} simulation. Figure~\ref{fig:cl_wmap7_mean} presents the mean angular power spectra  derived from these  realizations for both cases across a range of frequencies, each corresponding to a different redshift bin. As shown, residuals typically range between $5\% -- 10\%$, again providing compelling evidence for the validity and accuracy of the remapping in the 21cm context.  The same results are found for the CPL and Planck18 cosmologies. These results are showcased in Figures Fig.~\ref{figure:cls_m003} for CPL and Planck18.

The results in this section present, for the first time, predictions for the 21cm signal derived from remapped DMHC. Although we applied this approach to three cosmologies (WMAP7, Planck18, and CPL), it can be extended to other cosmologies, offering insights into the angular clustering of the 21cm signal. The evolution of $C_\ell$ with redshift provides valuable information for future 21cm intensity mapping surveys to help extract information from upcoming observational data.

%%%%%%%%%%%%%%%%%%
\section{Conclusions and final remarks}
%%%%%%%%%%%%%%%%%%%
In this study, we explored the possibility of applying the remapping technique to DMHC for the generation of 21cm mocks. First, we validated the remapping process by comparing the halo and matter power spectra for three different cosmologies: WMAP7, Planck18, and CPL, all starting from HR4 N-body simulations based on WMAP5 cosmology. Our results indicate a $5-10\%$ residual between remapped and N-body products, a residual we consider sufficient for our purposes, especially in the range of scales where BINGO will operate. Using these remapped halo catalogs and the HOD method, we constructed 21cm intensity maps in frequency and redshift ranges compatible with those of the BINGO project. A comparison of these mocks, based on their angular power spectra, also shows a $5-10\%$ residual.\\

The results presented in this work are, to the best of our knowledge, the first time in which remapping can be extended to the construction of 21cm mocks in different cosmologies. This opens the possibility of a systematic simulation of  21cm intensity map signal, which is necessary for survey design and data analysis. We expect that the introduction of the additional steps suggested by \citep{Mead2014a} and a more careful treatment of the bias considering its mass dependence can improve the accuracy. Notably, our initial investigation into the debiased halo power spectra from the remapped catalogs suggests that standard halo bias prescriptions, while providing a good starting point, might benefit from further refinement when applied to remapped simulations. 

We observed that subtle adjustments to the bias values could lead to a better agreement with the theoretical matter power spectra, potentially reducing the need for stringent filtering of the halo catalogs. This indicates that the remapping process might introduce subtle alterations to halo properties that affect their clustering, suggesting the potential for developing bias calibrations specifically tailored for remapped catalogs in future work. However, as pointed out previously, the combination of the precision achieved by our methodology and the computational efficiency of the method applied here is sufficiently good to be considered optimal for our current goals.

A key aspect of our study, detailed in Appendix~\ref{Remapping precision}, offers a quantified analysis of the remapping precision by examining the impact of cosmological parameter variations on the rescaling factors $s$ and $s_m$. Our findings establish clear thresholds for the rescaling parameters to achieve accurate remapping ($s \lesssim 10\%$ and $s_m \lesssim 25\%$) and identify regimes where remapping becomes unreliable (e.g., residuals $> 50\%$ when $s > 20\%-30\%$ or $s_m > 30\%-50\%$). Notably, a large $s_m$ was found to be a strong indicator of poor remapping accuracy. This systematic study provides valuable guidance on the relevance and limitations of the remapping technique for various cosmological transitions.

The efficiency of the remapping technique offers a significant advantage over computationally expensive direct N-body simulations, particularly when exploring a range of cosmological parameters or generating large numbers of mock observations required for survey calibration and error analysis. While we have focused on the large-scale clustering and the 21cm intensity mapping signal, the remapped halo catalogs can be further utilized for various downstream applications, such as galaxy formation modeling and the generation of galaxy survey mocks. Ultimately, the remapping technique presented here provides a powerful tool for efficiently bridging the gap between theoretical cosmology and observational probes of large-scale structure, especially in the context of upcoming 21cm intensity mapping experiments aiming to unravel the mysteries of dark energy and the evolution of the Universe.

%%%%%%%%%%%%%%%%%%%%%%%%%%%%%%%%%%%%%%%%%%%%%%%%

%%%%%%%%%%%%%%%%%%%%%%%%%%%
\section*{Acknowledgements}

The authors acknowledge the use of  data from the Horizon Run4~\citep{Kim2015}.We also acknowledge the use of the CHE
cluster, managed and funded by the COSMO/CBPF/MCTI,
with financial support from FINEP and FAPERJ, operating
at Javier Magnin Computing Center/CBPF. Additionally, we acknowledge the use of the healpy/HEALPix package~\citep{Gorski2005} for processing and analyzing data. This research also made use of the \texttt{nbodykit} toolkit~\citep{Hand2018} for large-scale structure analysis.RM acknowledges the financial support from CNPq under the fellowship Processo 302370/2024-2. BBB acknowledges the financial support from FAPESP under the fellowships 2022/16749-0 and 2024/12902-3. WSHR acknowledges support from FAPES. C. A. Wuensche thanks CNPq for grants 312505/2022-1 and 407446/2021-4. R. Mokeddem thanks G. A. S. Silva for useful comments and the reading of the manuscript. C.A. Wuensche thanks CNPq for grants 407006/2021-4 and 312505/2022-1, the Brazilian Ministery of Sciencem Technology and Innovation (MCTI) and the Brazilian Space Agency (AEB) who supported the present workd under the PO20VB.0009.

%%%%%%%%%%%%%%%%% APPENDICES %%%%%%%%%%%%%%%%%%%%%

\appendix

\section{Remapping Precision and Limitations} \label{Remapping precision}

This appendix explores the precision of the remapping process by examining how variations in cosmological parameters influence the rescaling factors, $s$ and $s_m$. Specifically, we analyze the impact of deviations of these factors from unity on the accuracy of the remapping. The central aim is to establish a quantitative relationship between the rescaling parameters and the residuals observed when comparing simulated matter power spectra to their theoretical counterparts. This analysis allows us to define the conditions under which remapping yields reliable results and to delineate its inherent limitations.

To systematically investigate this, we constructed a suite of target cosmologies by varying the parameters around three base models: WMAP7, M003, and Planck2018. The resulting cosmologies are labeled to reflect their origin (e.g., "Planck18-like 2"). Table~\ref{tab:cosmo_param} summarizes the cosmological parameters of these models, along with the computed values of the rescaling factors $s$ and $s_m$, and the corresponding redshifts $z$ and $z'$ derived from the first step of the remapping process.

\begin{figure}
    \centering
    \includegraphics[height=16.5cm, width=16.5cm]{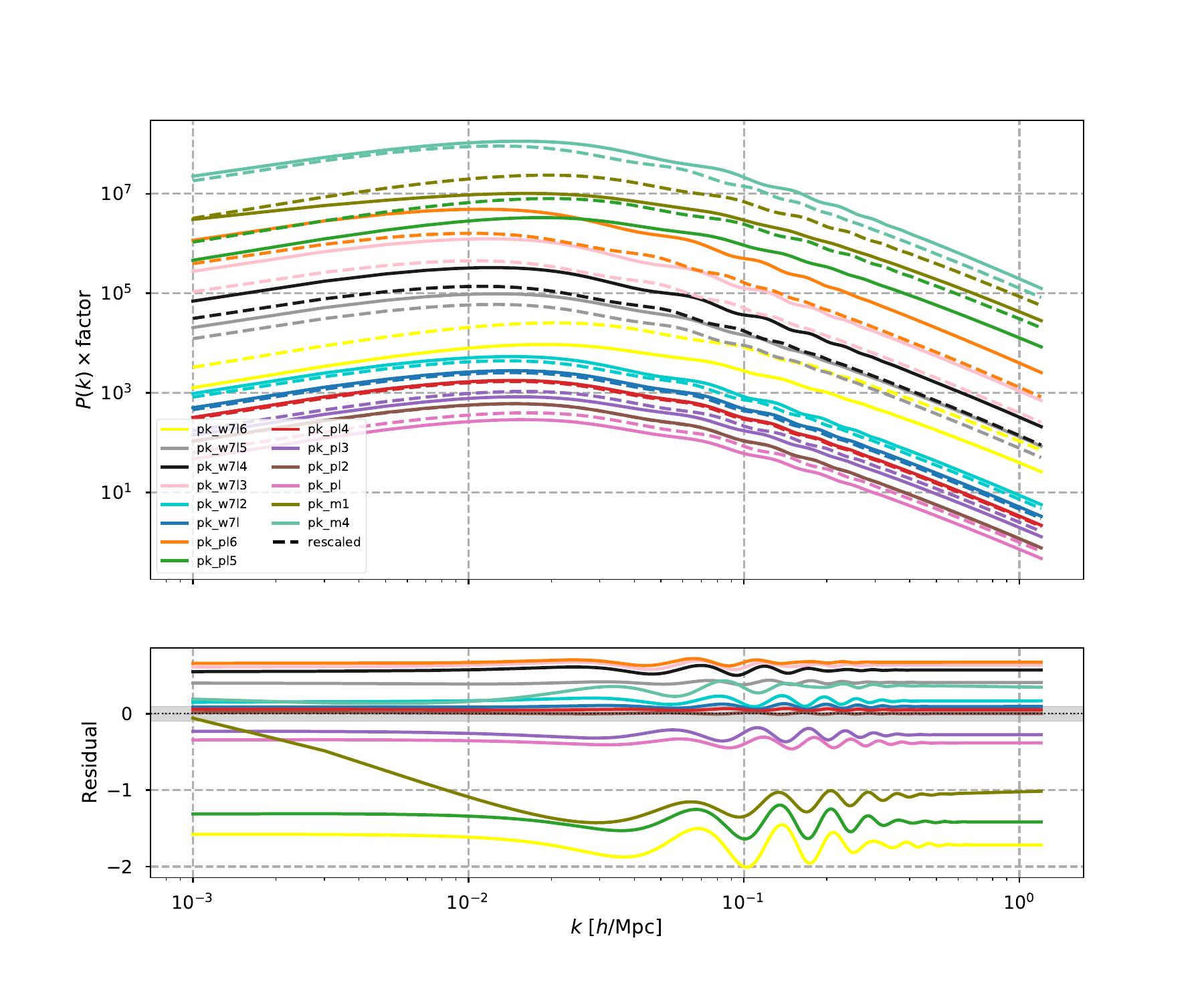}
    \caption{Top: theoretical matter P(k) of cosmologies from Table~\ref{tab:cosmo_param} (multiplied by separation factors to make the plot clear). The continued lines are of the target cosmology, the dashed lines are the rescaled WMAP5 matter power spectra. Bottom: this plot illustrates the accuracy of the remapping process across various cosmological models, showing how the residuals, representing remapping quality, can sometimes fall outside the accepted region of $\pm10\%$.}
    \label{fig:all_remapped_theo}
\end{figure}

\begin{figure}
    \centering
    \includegraphics[height=13cm, width=13cm]{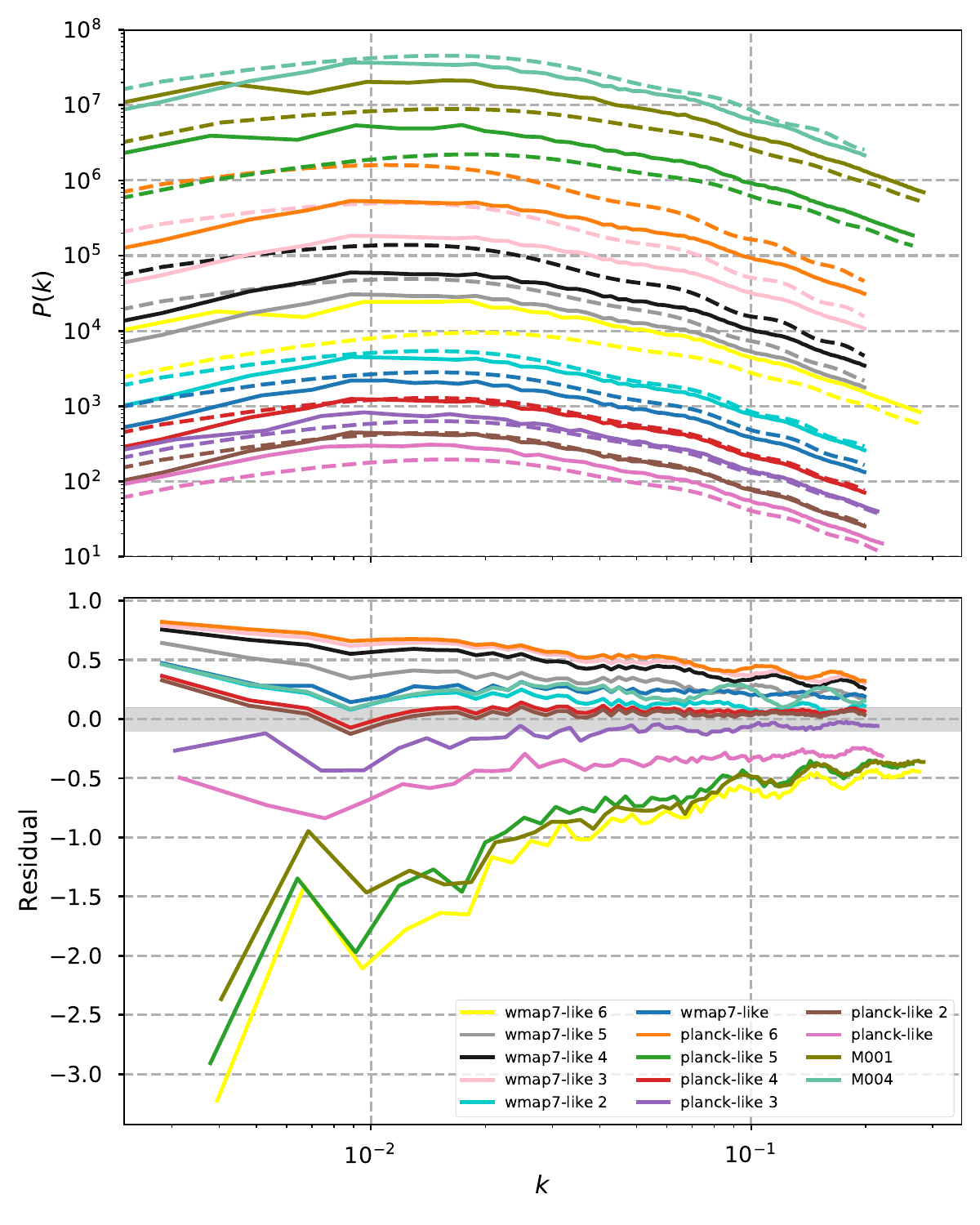}
    \caption{%Matter P(k) theoretical and from simulations of Table~\ref{tab:cosmo_param} (multiplied by separation factors to make the plot clear). 
    Top: Theoretical and simulated (from Table~\ref{tab:cosmo_param}) power spectra P(k), multiplied by separation factors to make the plot clear. Bottom: residuals of upper plot simulations. This figure demonstrates the complete remapping process, showing the matter power spectra from simulations. It validates the insights from Figure~\ref{fig:all_remapped_theo}, confirming that the remapping performance can indeed be predicted based on the $s$ and $s_m$ parameters.}
    \label{fig:all_remapped}
\end{figure}

Initially, we conducted a comparative analysis of the theoretical matter power spectra as shown in Figure~\ref{fig:all_remapped_theo} to identify scenarios where significant residuals suggest potential limitations of the remapping technique. Subsequently, we applied the remapping procedure to transform the original simulation into each of the target cosmologies. The remapping accuracy was then evaluated by comparing the power spectra of the remapped simulations with the theoretical predictions, as shown in Figure~\ref{fig:all_remapped}.

\begin{table*}[h!]
    \centering % Use \centering instead of \begin{center}...\end{center}
    \caption{Cosmological Parameters and Remapping Factors. This table presents the cosmological parameters for several custom cosmological models, along with their corresponding $z, s$ and $s_m$ remapping factors. it is used for further investigations in Appendix~\ref{Remapping precision}.} 
    \resizebox{\textwidth}{!}{
    \begin{tabular}{|l|c|c|c|c|c|c|c|c|c|c|c|} % Removed unnecessary vertical lines for cleaner look
        \hline 
        Cosmology/Model & $h$ & $\Omega_m$ & $\Omega_b$ & $\sigma_8$ & $n_s$ & $\omega_0$ & $\omega_a$ & $z$ & $z'$ & $s$ & $s_m$ \\
        \hline \hline
        WMAP5 & 0.72 & 0.26 & 0.043 & 0.793 & 0.96 & -1.0 & 0 & $-$ & $-$ & $-$ & $-$ \\
        \hline
        WMAP7 & 0.71 & 0.266 & 0.044 & 0.8 & 0.963 & -1.0 & 0 & 0.4 & 0.43 & 0.9873 & 0.9845 \\
        WMAP7-like & 0.67 & 0.26598 & 0.044 & 0.8 & 0.963 & -1.0 & 0 & 0.15 & 0.12 & 1.033674 & 1.1298 \\
        WMAP7-like 2 & 0.67 & 0.266 & 0.0503 & 0.8 & 0.963 & -1.0 & 0 & 0.3 & 0.22 & 1.075 & 1.271 \\
        WMAP7-like 3 & 0.67 & 0.2099 & 0.044 & 0.8 & 0.963 & -1.0 & 0 & 0.51 & 0.1 & 1.4034 & 2.231 \\
        WMAP7-like 4 & 0.71 & 0.2099 & 0.044 & 0.8 & 0.963 & -1.0 & 0 & 0.4 & 0.05 & 1.3268 & 1.885 \\
        WMAP7-like 5 & 0.71 & 0.2099 & 0.03 & 0.8 & 0.963 & -1.0 & 0 & 0.4051 & 0.21 & 1.1918 & 1.3667 \\
        WMAP7-like 6 & 0.71 & 0.35 & 0.05 & 0.8 & 0.963 & -1.0 & 0 & 0.15056 & 0.57 & 0.7166 & 0.4955 \\
        \hline
        CPL(M003) & 0.7167 & 0.3017 & 0.0427 & 0.9 & 0.8944 & -1.10 & -0.2833 & 0.153 & 0.401 & 0.9874 & 1.117 \\
        CPL M001 & 0.6167 & 0.387 & 0.0366 & 0.8778 & 0.9611 & -0.70 & 0.6722 & 0.00088 & 0.5 & 0.7635 & 0.66268 \\
        CPL M004 & 0.5833 & 0.3641 & 0.06709 & 0.7889 & 0.8722 & -1.1670 & 1.1500 & 0.50262 & 0.25 & 1.1565 & 2.16712 \\
        \hline
        Planck18 & 0.676 & 0.309 & 0.0489 & 0.8102 & 0.9665 & -1.0 & 0 & 0.291 & 0.502 & 0.8616 & 0.7601 \\
        Planck18-like & 0.71 & 0.28266 & 0.0444 & 0.8102 & 0.9665 & -1.0 & 0 & 0.108 & 0.29 & 0.8995 & 0.7908 \\
        Planck18-like 2 & 0.71 & 0.2533 & 0.03979 & 0.8102 & 0.9665 & -1.0 & 0 & 0.20344 & 0.25 & 1.0026 & 0.98194 \\
        Planck18-like 3 & 0.75 & 0.2533 & 0.03565 & 0.8102 & 0.9665 & -1.0 & 0 & 0.1 & 0.26 & 0.92444 & 0.76965 \\
        Planck18-like 4 & 0.676 & 0.26 & 0.04 & 0.8102 & 0.9665 & -1.0 & 0 & 0.20 & 0.219 & 1.0187 & 1.05722 \\
        Planck18-like 5 & 0.676 & 0.35 & 0.05 & 0.8102 & 0.9665 & -1.0 & 0 & 0.1054 & 0.5 & 0.7464 & 0.5598 \\
        Planck18-like 6 & 0.676 & 0.19 & 0.035 & 0.8102 & 0.9665 & -1.0 & 0 & 0.51 & 0.08 & 1.455 & 2.2537 \\
        \hline
    \end{tabular}
    }
    \label{tab:cosmo_param} 
\end{table*}

Based on this analysis, we can categorize the remapping outcomes into two primary scenarios:

\begin{enumerate}
    \item \textbf{Accurate Remapping:}  When the residuals between the simulated and theoretical matter power spectra are below $20\%$, the remapping is considered reasonably accurate. In these cases, further refinement can be achieved by applying additional thresholds on the simulation box size and/or halo mass, which typically reduces the residuals to below the accepted threshold of $10\%$.
    \item \textbf{Inaccurate Remapping:}  When the residuals exceed $20\%$, the remapping technique is deemed unreliable for accurately simulating the specific target cosmology.
\end{enumerate}

These findings provide crucial insights into the limitations of the remapping method and offer practical guidance on its appropriate application. The following subsections detail specific thresholds and regimes that influence remapping accuracy.

\subsection{Optimal Range for Accurate Remapping}

Accurate remapping, characterized by minimal residuals, is generally achieved when the rescaling parameters are within the following ranges:

\begin{itemize}
    \item $s \lesssim 1.1$  \quad (i.e., deviations of less than $10\%$ from unity)
    \item $s_m \lesssim 1.25$ \quad (i.e., deviations of less than $25\%$ from unity)
\end{itemize}

\subsection{Threshold for Increased Residuals \texorpdfstring{$> 50\%$}{> 50\%}}

As the rescaling parameters deviate further from unity, the accuracy of the remapping deteriorates:

\begin{itemize}
    \item When $s$ exceeds approximately 1.2 to 1.3 ($20\%-30\%$ deviation), residuals begin to increase noticeably.
    \item When $s_m$ exceeds approximately 1.3 to 1.5 ($30\%-50\%$ deviation), inaccuracies become significant, often resulting in residuals greater than $50\%$.
\end{itemize}

\subsection{Critical Regime for Severe Inaccuracies \texorpdfstring{$> 100\%$}{> 100\% residuals}}

In a more extreme regime, the remapping becomes highly unreliable:

\begin{itemize}
    \item Values of $s$ around 1.25 to 1.28 and $s_m$ around 1.33 to 1.5 correspond to residuals that can reach up to $150\%$, indicating a breakdown of the remapping process.
\end{itemize}

\subsection{Dominant Effect of Large \texorpdfstring{$s_m$}{sm} Values}

Our analysis reveals that the mass rescaling factor, $s_m$, plays a particularly important role in determining the accuracy of the remapping:

\begin{itemize}
    \item  Very large values of $s_m$ (e.g., $> 1.8$ or $80\%$ deviation) tend to be associated with substantial residuals ($40\%-60\%$), even when $s$ is moderately within the acceptable range (e.g., up to 1.3 or $30\%$ deviation).
    \item This suggests that while both $s$ and $s_m$ influence the residual, deviations in $s_m$ are a stronger indicator of potential inaccuracies in the remapping.
\end{itemize}

\subsection{General Guideline for Accurate Remapping}

In summary, for accurate remapping with minimal residuals, it is crucial to maintain $s$ below approximately 1.1 and $s_m$ below 1.25. Exceeding these thresholds, especially when $s$ surpasses 1.2-1.3 or $s_m$ surpasses 1.3-1.5, leads to a rapid increase in residuals, potentially exceeding $50\%$ and reaching up to $150\%$ in extreme cases.

\begin{table*} %[h!]
    \centering
    \caption{Deviations of \( s \) and \( s_m \) from 1 for different cosmological models.}
    \label{tab:s_deviation}
    \begin{tabular}{|l|c|c|c|c|c|c|}
        \hline
        \textbf{Cosmology} & \( s \) & \( s_m \) & \( |\Delta s| \) & \( |\Delta s_m| \) & \( |\Delta s| (\%) \) & \( |\Delta s_m| (\%) \) \\
        \hline
        WMAP7 & 0.9873 & 0.9845 & 0.0127 & 0.0155 & 1.27 & 1.55 \\
        WMAP7 like & 1.0337 & 1.1298 & 0.0337 & 0.1298 & 3.37 & 12.98 \\
        WMAP7 like 2 & 1.0659 & 1.2389 & 0.0659 & 0.2389 & 6.59 & 23.89 \\
        WMAP7 like 3 & 1.4034 & 2.2310 & 0.4034 & 1.2310 & 40.34 & 123.10 \\
        WMAP7 like 4 & 1.3268 & 1.8850 & 0.3268 & 0.8850 & 32.68 & 88.50 \\
        WMAP7 like 5 & 1.1918 & 1.3667 & 0.1918 & 0.3667 & 19.18 & 36.67 \\
        WMAP7 like 6 & 0.7166 & 0.4955 & 0.2834 & 0.5045 & 28.34 & 50.45 \\
        CPL M003 & 0.9874 & 1.1170 & 0.0126 & 0.1170 & 1.26 & 11.70 \\
        CPL M001 & 0.7635 & 0.6627 & 0.2365 & 0.3373 & 23.65 & 33.73 \\
        CPL M004 & 1.1565 & 2.1671 & 0.1565 & 1.1671 & 15.65 & 116.71 \\
        Planck & 0.8616 & 0.7601 & 0.1384 & 0.2399 & 13.84 & 23.99 \\
        Planck like & 0.8995 & 0.7908 & 0.1005 & 0.2092 & 10.05 & 20.92 \\
        Planck like 2 & 1.0026 & 0.9819 & 0.0026 & 0.0181 & 0.26 & 1.81 \\
        Planck like 3 & 0.9244 & 0.7697 & 0.0756 & 0.2304 & 7.56 & 23.04 \\
        Planck like 4 & 1.0187 & 1.0572 & 0.0187 & 0.0572 & 1.87 & 5.72 \\
        Planck like 5 & 0.7464 & 0.5598 & 0.2536 & 0.4402 & 25.36 & 44.02 \\
        Planck like 6 & 1.4550 & 2.2537 & 0.4550 & 1.2537 & 45.50 & 125.37 \\
        \hline
    \end{tabular}
\end{table*}

%%%%%%%%%%%%%%%%%%%%%%%% pk comparison %%%%%%%%%%%%%%%%%%%%%%%%%%%%%%%%

\begin{figure}
    \centering
    \includegraphics[height=13cm, width=13cm]{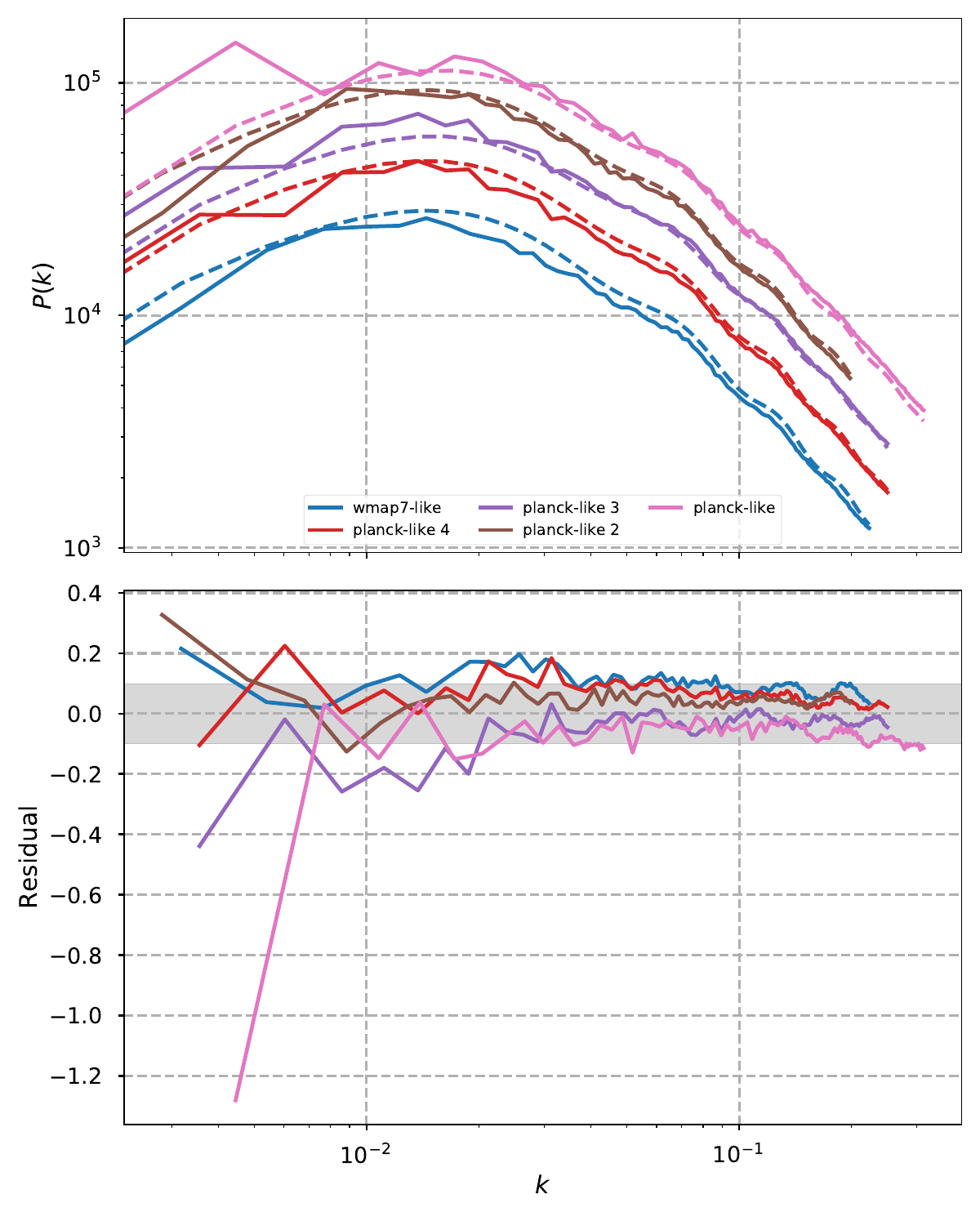}
    \caption{Matter P(k) theoretical and from simulations of Table~\ref{tab:cosmo_param} where the residual less than $10\%$}
    \label{fig:pk_appendix1}
\end{figure}

\begin{figure}
    \centering
    \includegraphics[height=13cm, width=13cm]{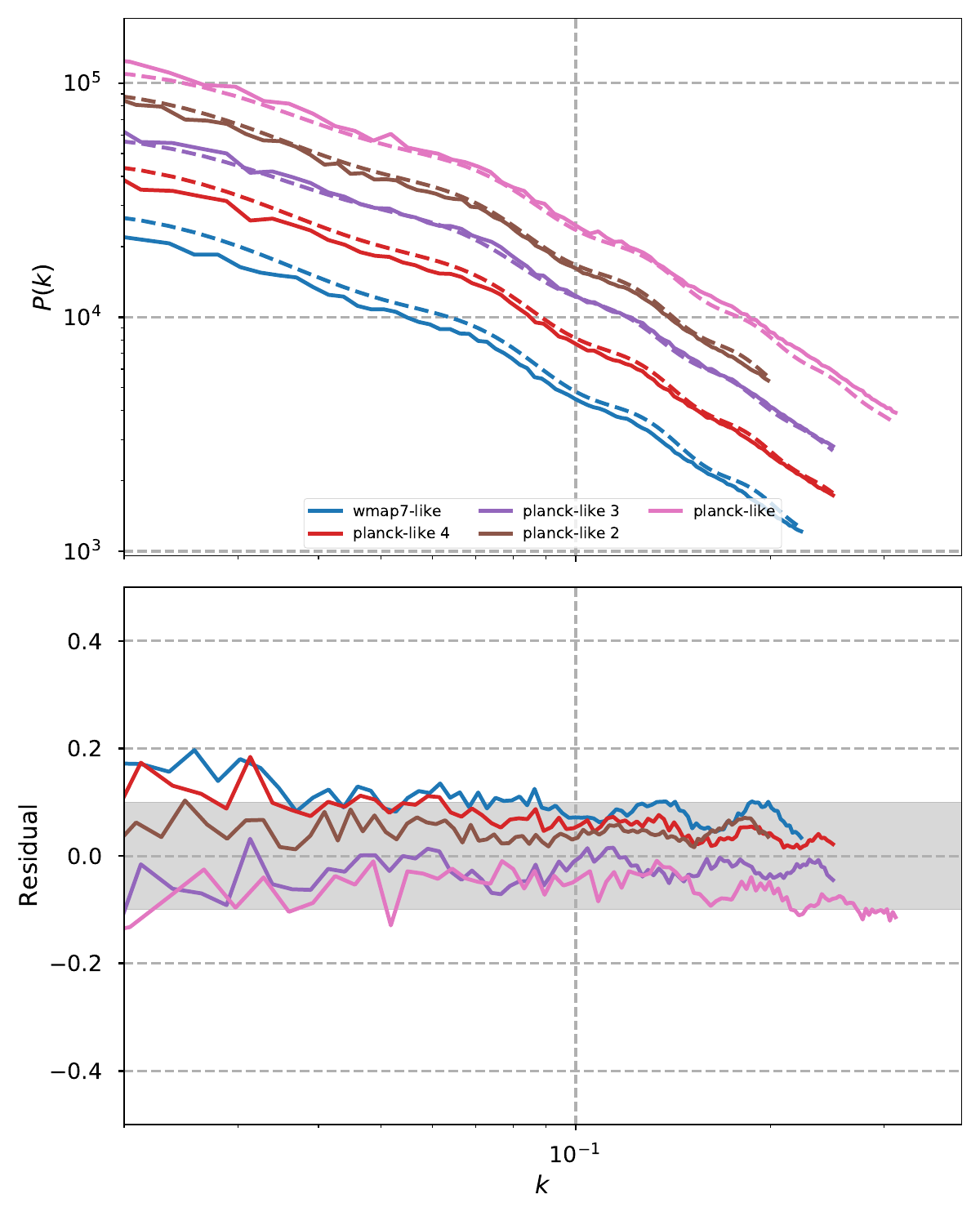}
    \caption{zoom in on Matter P(k) theoretical and from simulations of Table~\ref{tab:cosmo_param} where a mass and a box threshold are needed}
    \label{fig:pk_appendix2}
\end{figure}
%%%%%%%%%%%%%%%%%%%%%%%%%%%%%% Appendix B %%%%%%%%%%%%%%%%%%%%%%%%%%%%%%%%%%

%%%%%%%%%%%%%%%%%%%% REFERENCES %%%%%%%%%%%%%%%%%%

% \begin{thebibliography}{99}

\end{document}